\documentclass[10pt,journal,compsoc]{IEEEtran}
\usepackage{amsmath,amsfonts}
\usepackage{algorithmic}
\usepackage{algorithm}
\usepackage{array}
\usepackage[caption=false,font=normalsize,labelfont=sf,textfont=sf]{subfig}
\usepackage{textcomp}
\usepackage{stfloats}
\usepackage{url}
\usepackage{verbatim}
\usepackage{graphicx}
\usepackage{cite}
\begin{document}

\title{Interactive Visual Analysis of Spatial Sensitivities}

\author{Marina~Evers, Simon~Leistikow, Hennes~Rave, and Lars~Linsen
\thanks{All authors are with the University of Münster, Germany. \protect\\
E-mail: \{marina.evers\textbar simon.leistikow\textbar hennes.rave\textbar linsen\}@uni-muenster.de.}
\thanks{Manuscript received ...; revised ...}}

\markboth{IEEE Transactions on Visualization and Computer Graphics}%
{Evers \MakeLowercase{\textit{et al.}}: Interactive Visual Analysis of Spatial Sensitivities}

\def\mycopyrightnotice{%
  \begin{minipage}{\textwidth}
  \centering \scriptsize
  Copyright~\copyright~2024 IEEE. Personal use of this material is permitted. Permission from IEEE must be obtained for all other uses, in any current or future media, including\\reprinting/republishing this material for advertising or promotional purposes, creating new collective works, for resale or redistribution to servers or lists, or reuse of any copyrighted component of this work in other works by sending a request to pubs-permissions@ieee.org.\\
	DOI: \url{10.1109/TVCG.2024.3433001}
  \end{minipage}
}
\IEEEpubid{\mycopyrightnotice}

\maketitle

\IEEEpubidadjcol

\begin{abstract}
Sensitivity analyses of simulation ensembles determine how simulation parameters influence the simulation's outcome. Commonly, one global numerical sensitivity value is computed per simulation parameter. However, when considering 3D spatial simulations, the analysis of localized sensitivities in different spatial regions is of importance in many applications. For analyzing the spatial variation of parameter sensitivity, one needs to compute a spatial sensitivity scalar field per simulation parameter. Given $n$ simulation parameters, we obtain multi-field data consisting of $n$ scalar fields when considering all simulation parameters. We propose an interactive visual analytics solution to analyze the multi-field sensitivity data. It supports the investigation of how strongly and in what way individual parameters influence the simulation outcome, in which spatial regions this is happening, and what the interplay of the simulation parameters is. Its central component is an overview visualization of all sensitivity fields that avoids 3D occlusions by linearizing the data using an adapted scheme of data-driven space-filling curves. The spatial sensitivity values are visualized in a combination of a Horizon Graph and a line chart. We validate our approach by applying it to synthetic and real-world ensemble data.
\end{abstract}


\section{Introduction}
\label{sec:introduction}
\IEEEPARstart{S}{imulation ensembles} of 3D spatial data are commonly generated in different areas of science like physics or geoscience as well as in medicine. Often, the modeled phenomena depend on a set of input parameters. As the exact parameter values are unknown or uncertain, they are varied and an ensemble is created where each simulation run corresponds to the outcome with respect to one parameter setting. 
As the simulations are often computationally expensive, it is desired to limit the number of parameters to the most important ones. \textit{Global sensitivity analysis} investigates how much the output's variation can be attributed to different input alterations and as such is a useful technique towards the goal of detecting which input parameters influence the simulation outcome most.

Global sensitivity analyses typically compute one sensitivity value per parameter. For spatial simulation data, however, it would be desirable to investigate, which spatial regions are more sensitive than others.
For example, when considering the outcome of a medical treatment simulation, it is of utmost importance to understand in which regions or on which tissues the simulation models are more sensitive:
In regions close to risk structures, less sensitivity is allowed to assure that the treatment is successful without harming any essential healthy tissues.
Such a \textit{spatial sensitivity analysis} is a little explored topic. 
One reason for the lack of spatial sensitivity analysis methods are challenges in the visualization of the spatial sensitivities. For simulations over a 3D spatial domain, the outcome of a global spatial sensitivity analysis is a multi-field volume data set, which contains one scalar field of sensitivities per input parameter. Given $n$ input parameters, this results in $n$ 3D scalar fields that need to be analyzed.

In this paper, we propose an approach to analyze multi-field sensitivity data that covers the spatial variation of the sensitivity.
For the computation of spatial sensitivities, we compare three different methods, namely Sobol indices~\cite{sobol2001global}, the $\delta$ sensitivity measures~\cite{borgonovo2007new}, and distance-based generalized sensitivity analysis~\cite{fenwick2014quantifying}. However, our visual analysis approach can also be used on other sensitivity methods that can be applied to each spatial sample separately. In particular, it is also applicable to more complex approaches involving surrogate models to reduce the computational complexity of the measure.

To analyze the multi-field sensitivity data, we propose an overview visualization of all sensitivity fields based on the combination of line charts and a Horizon Graph that uses nested, superimposed bands for space-efficient visualization.
For its layout, we propose to use a projection of the 3D spatial data to a 1D embedding using the concept of a space-filling curve, which allows for the simultaneous investigation of all spatial sensitivities without the 3D occlusion problems inherent to volume visualization.
To preserve homogeneous sensitivity regions during projection, i.e., regions in which the same parameters are sensitive, we adapt the objective function of data-driven space-filling curves~\cite{zhou2020data}, see Section~\ref{sec:sensitivityVis}.

In our overview visualization, we can identify high-sensitivity spatial regions as well as high-sensitivity parameters, which can then be analyzed in more detail.
In particular, when having detected a parameter with high sensitivity, one is interested in examining how the simulation outcome depends on this parameter. A linked view showing the simulation output with respect to the chosen parameter's values and the spatial locations supports such an analysis, see Section~\ref{sec:parameterDependency}.
These views are complemented by a parallel coordinates plot that allows for the analysis of sensitivity values of individual parameters as well as their correlation, see Section~\ref{sec:pcp}, and a surface rendering of brushed spatial regions in 3D space. 

The problem specification and the targeted analysis tasks are detailed in Section~\ref{sec:problem_specification}, while Section~\ref{sec:overview} presents an overview of the components of our visual analytics solution.

Our main contributions can be summarized as follows:
\begin{itemize}
    \item An overview visualization of the multi-field sensitivity data. It is based on an adapted scheme for data-driven space-filling curves to allow for a global overview of all spatial locations without 3D occlusion issues of volume visualizations.
    \item An interactive visual analytics solution that allows for a comprehensive analysis of the spatial sensitivities. It supports the analysis of quantitative and qualitative influence of individual parameters on the simulation outcome, the corresponding spatial regions, and the relation between the parameters.
    \item An evaluation of the space-filling curve and sensitivity computation algorithms from which we derive guidelines on which algorithms to choose. We also apply our approach to two real-world medical simulations.
\end{itemize}

\section{Related work}
Recently, the analysis of spatio-temporal ensemble data has gained major attention~\cite{Wang2019, Crossno2018, Kehrer2013, wilson2009toward}. Several works focus on the uncertainty that is captured within the ensemble due to choosing different initial conditions or on the visualization of aggregated data such as means~\cite{potter2009ensemble, sanyal2010noodles}.
Another common task is the parameter-space analysis of the parameters that form the input of the ensemble. Several visualization methods including parallel coordinate plots, radial plots, scatter plots, line charts, matrices, and glyphs have been proposed~\cite{obermaier2015visual, bruckner2010result, fofonov2018projected, unger2012visual, Wang2017, Luboschik2014, Orban2019, bock2015visual}. Sedlmair et al.~\cite{sedlmair2014visual} proposed a comprehensive framework that guides the development of further parameter-space analysis research. They identified \textit{sensitivity analysis} as one of the core tasks. ParaGlide~\cite{bergner2013paraglide} is an interactive visualization system for the exploration of parameter spaces including the sensitivity of parameters, but mainly focuses on the stability of the results. Partitioning the parameter space and visualizing the partitioning using hyper-slices provides insights into the relation between the input parameters and the spatio-temporal simulation output~\cite{evers2022multi}. Kumpf et al.~\cite{kumpf2021visual} address the visualization of multi-field data and use brushing together with a visualization of field distributions to analyze ensembles, but do not include a sensitivity analysis.

Sensitivity analysis of ensemble data is a common task that can be divided into local sensitivity analysis, where the influence of small changes of one parameter is investigated, and global sensitivity analysis that takes the whole parameter space into account~\cite{saltelli2008global}. A good overview of sensitivity analysis techniques is provided by Pianosi et al.~\cite{pianosi2016sensitivity}. While our work focuses on the sensitivity of input parameters, ensemble sensitivity analysis (ESA) is a technique that studies the sensitivity on initial conditions instead of parameters~\cite{kumpf2018visual}.
Recently, some visual tools that provide tools for local sensitivity analysis have been proposed~\cite{Piringer2010, bergner2013paraglide, berger2011parameterspace, brecheisen2009parameter}.
Only few visual analysis tools are aiming at global sensitivity analysis. Fanovagraph~\cite{fruth2013fanovagraph} proposes a graph-based visualization of Sobol indices. This approach was extended by Yang et al.~\cite{yang2021senvis} to develop a computationally efficient tool for the analysis of Sobol indices of different order. Ballester-Ripoll et al.~\cite{ballester2019sobol, ballester2018tensor} use tensor-train models for the efficient computation of Sobol indices. However, they do not include the spatial variability in their approach but instead investigate the Sobol indices of scalar data. Directly applying their approach would require developing a tensor train surrogate for each spatial sample which is computationally very expensive.
Other approaches propose a more qualitative sensitivity analysis by showing the dependence of the outcome on single parameters~\cite{Luboschik2014, matkovic2009interactive}. InSituNet~\cite{He2019} is a surrogate model to analyze the parameter space. The authors directly use this model to derive the sensitivity of the output to the input parameters.

In conclusion, none of the discussed approaches take the spatial dimension of the data into account. With the increasing computing power available, \textit{spatial sensitivity analysis} has become feasible~\cite{ligmannZielinska2014spatially, xu2013spatially}. While most approaches show maps side-by-side, \c{S}alap-Ay\c{c}a et al. studied the use of visual stacking in comparison to adjacent maps. However, the coincident maps only hold for two-dimensional data and they rated the visualization of global spatial sensitivities as an open challenge where new methods are needed. 
Biswas et al.~\cite{biswas2016visualization} also tackle the challenge of spatial sensitivity visualization and even include the temporal domain. They use spatial clustering combined with maps and thus target their approach towards two-dimensional data, while our approach supports the analysis of 3D data and also considers the interaction between parameters.

\section{Problem specification}
\label{sec:problem_specification}
We consider a simulation ensemble that was created using a model with $n$ input parameters $p_i$, $i=1,...,n$. Here, we assume the input parameters to be numerical and continuous. In the following, whenever we use the short notation  ``parameters'', we refer to these simulation input parameters. Each simulation run $r$ of the ensemble is characterized by a set of $n$ input parameter values $(p_{1r},...,p_{nr})$. The output of a simulation run is a scalar field $S_r$ over a 2D or 3D spatial domain $D$, i.e., each spatial sample $\mathbf{x}\in D$ contains a scalar value $s_r(\mathbf{x})$. 

The main objective of the analysis we want to support with our visual analytics approach is to analyze the dependency of the simulation outcome on its input parameters. To achieve this main objective, we target the following individual tasks, which have already been identified in literature as being important for ensemble analysis~\cite{sedlmair2014visual, csalap2021less}:

\noindent
\textbf{(T1)} Determine the \textit{quantitative influence of individual parameters} on the simulation outcome, i.e., estimating how \textit{sensitive} the outcome is with respect to the parameter choices. For computational steering purposes, it is on the one hand desirable to identify the most influential parameters, while on the other hand it is also of interest to identify parameters with little influence. Then, when executing further simulation runs, one would sample the most influential parameters more densely, while not varying the settings for parameters with negligible influence.

\noindent
\textbf{(T2)} Analyze \textit{spatial sensitivities} of the outcome to parameters. 
Different spatial regions may be of varying importance for the simulation outcome, e.g., in case of risk structures in medical applications. Hence, the sensitivity analysis should be spatially resolved, i.e., the sensitivities should be computed for each spatial location $\mathbf{x}\in D$ individually.
Then, spatial regions of high/low sensitivities can be identified, further analyzed, and related to the simulated phenomenon.

\noindent
\textbf{(T3)} Investigate \textit{correlations of sensitivities} to different parameters. Observing the interplay of two parameters on the simulation output provides insights into how the underlying model, on which the simulations are based, functions. Again, this analysis should be spatially resolved to identify spatial regions of interest.

\noindent
\textbf{(T4)} Analyze \textit{the simulation outcome's qualitative dependency} on the input parameters.
Having identified spatial regions of interest, e.g., high-sensitivity regions with respect to some relevant parameter (see task T1), it is of interest to observe how that parameter affects the simulation output. This allows for, e.g., identifying increases or decreases over that parameter value.

\section{Interactive Spatial Sensitivity Analysis}
\label{sec:overview}

\begin{figure}
\centering
\includegraphics[width=\linewidth]{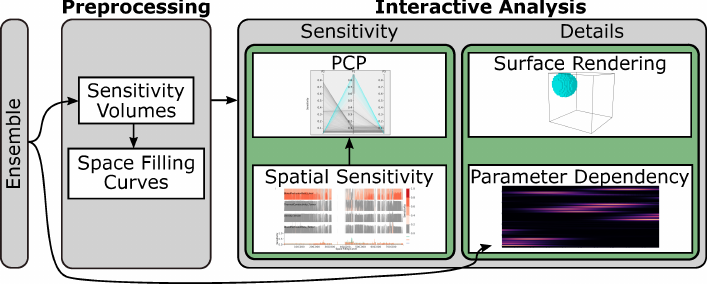}
\caption{Workflow of our approach: In a preprocessing step, the sensitivity volumes and the space-filling curve are calculated from the simulation ensemble. These data are used as input to the interactive analysis. The parallel coordinates plot (PCP) and spatial sensitivity visualization are linked and visualize the sensitivity volumes. Selections in those visualizations can be shown in more detail in the surface rendering and the parameter dependency visualization that shows the ensemble's simulation output.}
\label{fig:overview}
\end{figure}

Our approach for the analysis of spatial sensitivities involves multiple computational steps as well as different visual encodings to support the analytical tasks (T1-T4) within a visual analytics pipeline as presented in Figure~\ref{fig:overview}. Please refer to the supplemental video for a better understanding of the interactions.

To compute the dependency of the simulation output on the parameters (T1) and to allow for a spatially resolved analysis of the dependencies (T2), we compute \textit{ spatial sensitivities}. 
We investigate the use of three different sensitivity measures, see Section~\ref{sec:sensitivityComputation}.
As a result of this pre-computation step, we obtain $n$ sensitivity scalar fields. More precisely, we obtain one sensitivity scalar field for each of the $n$ parameters. 
Hence, in the subsequent steps, we need to analyze multi-field volume data consisting of $n$ scalar fields.

The first task (T1) is to assess the influence of individual parameters on the simulation outcome, i.e., we want to analyze the computed sensitivity values. 
As discussed in Section~\ref{sec:pcp}, we decided to use a 
\textit{parallel coordinates plot (PCP)} (see Figure~\ref{fig:teaser}), as it allows for the analysis of the sensitivity values in each of the dimensions (T1), scales well with the number of dimensions $n$, thus provides an overview of the sensitivity distributions, and also allows for the analysis of correlations between dimensions shown as adjacent axes (T3). Each PCP axis corresponds to the sensitivity values with respect to one input parameter (T1), which supports rating their influence.
Interactive reordering of the axes and brushing facilitates the analysis of correlations between sensitivities to different parameters (T3).

Having computed spatial sensitivities, we also want to analyze the spatial positions of the sensitivity values (T2).  
Brushing in the PCP allows for the selection of sensitivity ranges on individual axes and even combinations of ranges on multiple axes. We can link the brushing interaction with a view of the spatial domain. A \textit{surface rendering} (see Figure~\ref{fig:teaser}d) then shows all voxels that have sensitivity values within the brushed interval(s). To investigate the internal structure, the user can adapt the opacity of the rendering. An alternative to the surface rendering that shows the selected voxel would be a direct volume rendering. 
However, as we work with multi-field data, it is unclear which sensitivity value should be encoded.

The coordinated views allow for the observation of the locations of selected sensitivity values but do not yet allow for 
a global understanding of the distributions of spatial sensitivities (T2). 
To deal with the inherent occlusion problem of 3D volume visualizations and show all sensitivity fields at once, we propose a novel \textit{overview visualization of spatial sensitivities} as shown in Figure~\ref{fig:teaser}b, which is based on projecting the volume data to 1D using a (pre-computed) \textit{space-filling curve} (SFC), as discussed in Section~\ref{sec:sensitivityVis}. Our visual encoding combines Horizon Graphs with line chart renderings to create a scalable overview visualization of the multi-field sensitivity data over the 1D SFC. 

While sensitivities indicate how much the simulation outcome depends on the parameters, it is then of interest to analyze the qualitative dependencies of the simulation output on identified parameters (especially those with the highest sensitivities) in more detail. One is interested in investigating how the simulation outcome depends on the parameter (T4). For this purpose, we include a \textit{parameter dependency visualization} that visualizes the simulation outcome with respect to the values of the selected parameter and the spatial positions as shown in Figure~\ref{fig:teaser}c. We exploit the SFC again to display the information in a 2D heatmap as detailed in Section~\ref{sec:parameterDependency}. It allows the user to identify patterns in the simulation outcomes with respect to the selected parameter and space.

The different views we introduce are coordinated using \textit{brushing and linking interactions}. More precisely, brushing in the PCP determines the data shown in the parameter dependency visualization as well as the surface rendering. Brushing one or multiple regions in the spatial sensitivity visualization triggers a highlighting of the selection in the PCP and the parameter dependency visualization as well as the surface rendering. Even though the spatial sensitivity visualization already contains the spatial information explicitly, linking to the surface rendering is necessary to interpret the 3D locations and to distinguish features in the data from artifacts that may have been introduced by the linearization of the data.

Our visual analytics approach is implemented in a web-based application using Dash and Plotly~\cite{plotly} for the basic visualizations and interactions, D3~\cite{bostock2011d3} for our novel visualization designs, and vtk~\cite{schroeder1998visualization} for the volume visualizations. Our source code can be found at \url{https://github.com/marinaevers/spatial-sensitivity}.

\section{Sensitivity Computation}
\label{sec:sensitivityComputation}
A variety of different measures for computing the sensitivity to input parameters exists~\cite{saltelli2008global}. In the scope of this paper, we considered three different methods that result in a quantitative measure. However, the chosen sensitivity computation method can be easily exchanged for the remainder of the approach. We investigated the use of first-order Sobol indices~\cite{sobol2001global,saltelli2008global}, which are among the most popular global sensitivity measures. An alternative is the use of a global sensitivity measure called $\delta$, which also has been used by Biswas et al.~\cite{biswas2016visualization} for the visual investigation of sensitivity in weather ensembles. In contrast to Sobol indices, $\delta$ does not require a specific sampling scheme for an efficient computation. A third measure is distance-based generalized sensitivity analysis (DGSA) as proposed by Fenwick et al.~\cite{fenwick2014quantifying}. Compared to the other two measures, the generalization to datatypes other than scalar fields is straight forward as the sensitivity computation is based on distances. Detailed descriptions of these three measures are provided in the supplementary material.

To investigate the spatial variation of the sensitivity, the individual sensitivity is computed for each spatial sample. As all described measures result in one sensitivity value per input parameter, we obtain one volume of sensitivity values for each individual parameter. In the following, we refer to these volumes as \textit{sensitivity volumes}. Optionally, also higher-order interactions between parameters can be considered, leading to one additional volume per interaction. The additional volumes can be directly included in the analysis, in the same way as the sensitivity volumes for individual parameters. Thus, at least $n$ sensitivity volumes for $n$ parameters need to be analyzed together.

\section{Parallel coordinates plot}
\label{sec:pcp}

Assuming $n$ parameters, the computation of all spatial sensitivities for individual parameters and a selected sensitivity measure according to Section~\ref{sec:sensitivityComputation} leads to multi-field sensitivity data with $n$ scalar fields as explained in Section~\ref{sec:overview}. 
Having computed these $n$ sensitivity volumes, we next want to analyze them according to the tasks identified in Section~\ref{sec:problem_specification}. 

\begin{figure*}
  \centering
  \includegraphics[width=\linewidth]{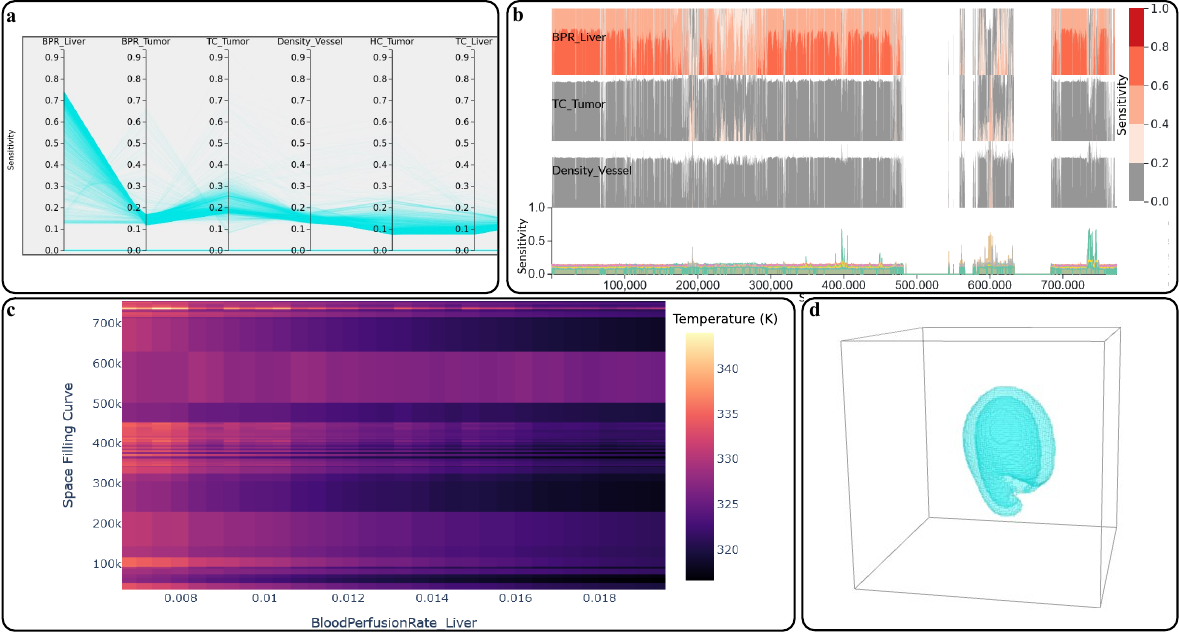}
  \caption{Interactive visual analysis of spatial sensitivities.
  Parallel coordinates (\textbf{a}) provide an overview of the sensitivities of input parameters and allow for brushing: A sensitivity visualization (\textbf{b}) shows the different sensitivity volumes over a 1D mapping using a data-driven space-filling curve. Both visualizations are linked to a surface rendering (\textbf{d}) that provides 3D spatial context. A parameter dependency visualization (\textbf{c}) supports an in-detail analysis of the parameters' influence on the simulation outcome.}
  \label{fig:teaser}
\end{figure*}

First, we want to investigate the \textit{influence of individual parameters}  on the simulation outcome (T1). Thus, we analyze the sensitivity value distribution for all parameters and interactions, which is a multidimensional data visualization task. 
Many multidimensional data visualization methods exist, but among the loss-less ones (excluding projections) the \textit{parallel coordinates plot} (PCP) scales best with the number of dimensions (better than table-based approaches or scatterplot matrices).
We set up the PCP by using one axis per sensitivity volume leading to $n$ axes.
Each voxel is then represented by a polygonal line in the PCP as shown in Figure~\ref{fig:teaser}a.
To allow interactive rates when brushing on PCP axes, we use a Monte Carlo subsampling to reduce the number of spatial sample points that are displayed. To improve the scalability with the potentially high number of axes, we introduce horizontal scrolling. We further sort the axes by the mean sensitivity of all voxels, i.e., the axes with highest sensitivities are shown first. A filtering operation allows the user to exclude irrelevant parameters from the further analysis by thresholding with a minimum percentage of sensitive voxels that is required to include the corresponding sensitivity volume in the analysis. 
By observing the whole PCP, we can directly compare the sensitivities on different parameters.
The axes of the plots are scaled equally to facilitate their comparison and avoid misinterpretations.

As an additional feature, our tool also supports including further domain-dependent information into the PCP, if available. An example would be the tissue types for the radiofrequency ablation simulation analyzed in Section~\ref{sec:usecases}. To visually differentiate the sensitivity axes from further information, we place the sensitivity axes in a labeled box, such that additional axes are rendered outside of this box.

The PCP further supports the analysis of \textit{correlations} of sensitivities (T3) by observing horizontal (positive correlation) or crossing (negative correlation) patterns between neighboring PCP axes, where the order of axes can be adjusted interactively.
To observe more subtle patterns, brushing on the PCP axes allows for selecting specific voxels based on their sensitivities.  

Having computed spatial sensitivities for all voxels, PCP allows for the analysis of the sensitivity values but omits the spatial information associated with each voxel.
Obviously, it is of high interest to observe where the voxels lie that have been detected to be highly sensitive (T2).
Therefore, we link the PCP to a \textit{surface renderer} (see Figure~\ref{fig:teaser}d), which visualizes the boundary surface of the intersection of all those voxels that contain sensitivity values within the intervals selected in the PCP.

\section{Calculation of space-filling curve}
\label{sec:sfc}
We have seen that the PCP presented in Section~\ref{sec:pcp} allows for a multi-dimensional analysis of the distributions of the sensitivity values. Brushing on the PCP axes and rendering the selected voxels in a surface renderer further allows for observing the 3D spatial positions of the selected voxels. However, it does not support the requirement of a global spatial overview (T2).
It is, for example, desired to detect regions, in which only one parameter is sensitive, several parameters are sensitive simultaneously, or the output is sensitive to none of the input parameters. While such regions can be identified in the PCP and rendered in the surface renderer, it requires a lot of interactions and cognitive effort to form a mental overview of the entire multi-field. 
To provide an overview of the multi-field sensitivity values in a spatial context, we use a space-filling curve (SFC) to create a 1D projection of the volumes such that the sensitivities can be shown as functions over that SFC. 
Our SFC-based visualization allows for a direct and immediate identification of the different regions.

Recently, several approaches using 1D projections to detect patterns in spatial data have been proposed~\cite{franke2021visual, buchmuller2018motionrugs, wulms2021stable} and applied to ensemble visualization. Demir et al.~\cite{demir2014multi} presented enhanced line charts for the analysis of ensemble data. Dynamic Volume Lines~\cite{weissenbock2018dynamic} use a nonlinear scaling of a Hilbert curve to support the analysis of volumetric ensemble data. Besides directly visualizing the data, brushing and linking to volume renderings are used~\cite{zhou2020data}. However, all of these approaches focus on visualizing the simulated ensemble volumes themselves with a special focus on the variations among the volumes. Thus, their visualizations are not optimized to show a selection of possibly rather different volumes on a joint domain, like our sensitivity volumes.

There exist various approaches for the design of SFCs, where Hilbert~\cite{hilbert1891ueber} and Peano curves~\cite{peano1890courbe} are among the most common ones. While they preserve locality well, they do not consider the underlying data. Additionally, its structure might lead to a splitting of features in the 1D projection, which was also observed in a study by Zhou et al.~\cite{zhou2020data}.
Therefore, we use an adaption of their \textit{data-driven space-filling curves}. This algorithm optimizes the curve to preserve both locality and features.

\begin{figure}
\centering
\includegraphics[width=\linewidth]{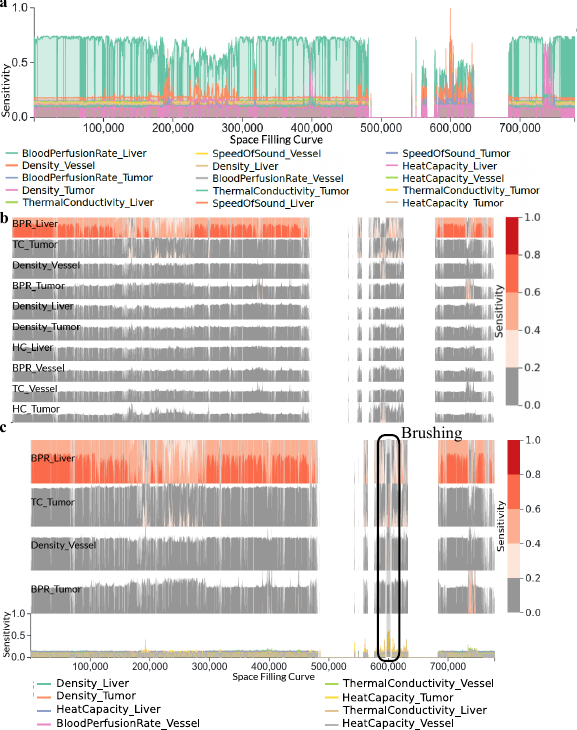}
\caption{Different visual encodings for the multi-field sensitivity data over a space-filing curve: While a line plot might cause overplotting (\textbf{a}), Horizon Graphs (\textbf{b}) become small with limited vertical space. Combining both visualizations (\textbf{c}) reduces these issues. Brushing (gray area) further links the plots vertically as well as to the other visualizations.}
\label{fig:sensitivityVisDesign}
\end{figure}

The calculation of data-driven SFC is based on the calculation of a Hamiltonian cycle. It is obtained by building a circuit graph, where adjacent grid points are combined to circuits. The dual graph of this circuit graph is used to determine a minimum spanning tree that is then used to calculate the Hamiltonian cycle. Zhou et al.~\cite{zhou2020data} proposed to use an objective function for weighting the dual graph, which reads
\begin{equation*}
    W(C_i, C_j) = (1-\alpha)N(C_i, C_j) + \alpha R(C_i, C_j).
\end{equation*}
Here, $C_i$ and $C_j$ are adjacent circuits and $\alpha$ is a user-defined blending factor. $N(C_i, C_j)$ is a term covering value coherency that was already used in context-based space filling curves~\cite{dafner2000context}. $R(C_i, C_j)$ covers positional coherency. The authors empirically found that $\alpha=0.1$ produces good results, which is the value we also use in our experiments. Both terms are normalized to the range $[0,1]$.

As the algorithm by Zhou et al.~\cite{zhou2020data} was developed to operate on a single scalar field, while we want to preserve features for multi-field data, we cannot directly apply the algorithm. If we had decided to choose only one of the sensitivity volumes for the SFC creation, we might have obtained a 1D projection that preserves well the features of that chosen single volume, while features of other volumes may not be preserved well. While we can keep the data-independent, positional coherency term as originally proposed, we modify the value coherency term to take all values of the multi-field data into account.

In its original version, the value coherency term is based on the magnitude of the differences between the data values~\cite{zhou2020data, dafner2000context}. 
Instead of using the differences between numbers $v_a$ and $v_b$, we 
investigate the following options for distances between two vectors $\boldsymbol{v}_a$ and $\boldsymbol{v}_b$, where the vectors are formed by the sensitivity values for the given parameters at the respective spatial positions: \\
\emph{$L_1$-norm}: Dafner et al.~\cite{dafner2000context} proposed to use the $L_1$-norm for a context-based SFC of RGB images. It is defined as 
\begin{equation*}
    d_1(\boldsymbol{v}_a, \boldsymbol{v}_b)=\sum_i|\boldsymbol{v}_{b,i}-\boldsymbol{v}_{a,i}|\ ,
\end{equation*}
where $\boldsymbol{v}_{a,i}$ and $\boldsymbol{v}_{b,i}$ correspond to the $i$-th component of $\boldsymbol{v}_a$ and $\boldsymbol{v}_b$, respectively. \\
\emph{$L_2$-norm/Euclidean distance}: One of the most common distance measures between two vectors is the Euclidean distance which can be computed as 
\begin{equation*}
    d_2(\boldsymbol{v}_a, \boldsymbol{v}_b)=\sqrt{\sum_i(\boldsymbol{v}_{b,i}-\boldsymbol{v}_{a,i})^2}\ .
\end{equation*}
\emph{$L_{\infty}$-norm}: The $L_{\infty}$-norm corresponds to the maximum value of two distances between the elements of the vector, given by 
\begin{equation*}
d_3(\boldsymbol{v}_a, \boldsymbol{v}_b)=\max_i(|\boldsymbol{v}_{b,i}-\boldsymbol{v}_{a,i}|)\ .
\end{equation*}
\emph{Sum of squared distances}: To save the computational costs for computing the square root of the Euclidean distances, we also consider the sum of squared distances 
\begin{equation*}
d_4(\boldsymbol{v}_a, \boldsymbol{v}_b)=\sum_i(\boldsymbol{v}_{b,i}-\boldsymbol{v}_{a,i})^2\ .
\end{equation*}
\emph{Cosine distance}: The cosine similarity is another common measure for investigating similarities between vectors and can be mapped to normalized distances by 
\begin{equation*}
d_5(\boldsymbol{v}_a, \boldsymbol{v}_b)=1-\frac{\sum_i \boldsymbol{v}_{a,i}\boldsymbol{v}_{b,i}}{\sqrt{\sum_i \boldsymbol{v}_{a,i}^2}\sqrt{\sum_i \boldsymbol{v}_{b,i}^2}}\ .
\end{equation*}
The distance measure $d_1$ to $d_4$ are normalized to the range $[0,1]$ while a normalization of $d_5$ is not necessary as all sensitivity values have to be non-negative. The original distance measure used by Zhou et al.~\cite{zhou2020data} is included as a limit case in the distance measures $d_1$ to $d_3$.

\section{Spatial sensitivity visualization}
\label{sec:sensitivityVis}

For the visualization of multi-field sensitivity data, we show the sensitivity data over the SFC. As the number of voxels is usually large, we use the same spatial subsampling scheme as applied in the PCP and only display the sensitivities for these spatial samples in the rendering. Note that the SFC is calculated on the complete volumes, but only the sensitivities for the respective subsampling positions are used for the plot. While Weissenböck et al.~\cite{weissenbock2018dynamic} use a nonlinear scaling to reduce the space needed for their visualization, we decide to use a linear one. In our case, it is not clear from the beginning what characterizes interesting regions that should be enhanced. Thus, a nonlinear scaling might visually reduce the impact of regions that would be rated as important otherwise. If the user identifies an interesting feature that appears small in the complete visualization, it is possible to zoom in and enlarge the corresponding interval of the SFC for a more detailed analysis.

For the \textit{visual encoding} of the sensitivity values, we considered several design alternatives. 
A first option would be to consider table-based visualizations such as a heatmap or a Table Lens, where each row represents one volume and each column one voxel in the SFC order. Reading off exact data values from heatmaps is difficult though. 
A Table Lens approach would be better in this regard, but does not scale to the number of voxels we want to visualize.
Two other options would be to display all sensitivity fields as continuous line plots or as discrete scatter plots in one coordinate system. 
Since we assume spatial coherence along the SFC, a continuous representation as in line plots is well justified. 
The discrete scatterplots, on the other hand, quickly produce visual clutter when rendering multiple fields.

So, our first design choice was to use \textit{line plots}.
However, line plots are difficult to interpret when rendering many lines due to overlap and significant fluctuations. Therefore, we color the areas below the lines. For a further reduction of overlap, we create a drawing order based on the number of sensitive voxels. Thus, the sensitivity values with more sensitive data are drawn in the back and the data showing smaller values in front. This also corresponds to the default ordering we used in the PCP, see Section~\ref{sec:pcp}. This visual design is shown in Figure~\ref{fig:sensitivityVisDesign}a.

\begin{figure}
\centering
\includegraphics[width=\linewidth]{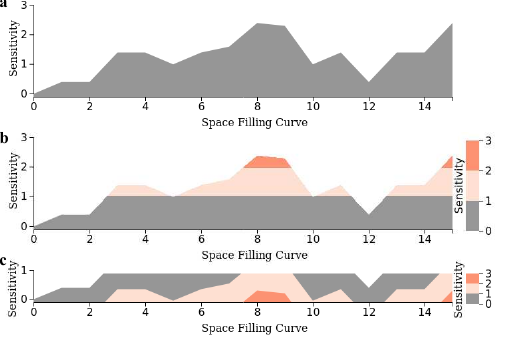}
\caption{Creation of Horizon Graphs for sensitivity visualization. After plotting an area chart (\textbf{a}), the plot is horizontally divided into bands, where each band is color-coded and has a height of $1$ (\textbf{b}) in case of DGSA and a height of $0.2$ for normalized sensitivity measures. Then, each band is moved to the baseline, leading to a constant height, independent of the occurring data values (\textbf{c}).}
\label{fig:horizonGraphCreation}
\end{figure}

Despite our efforts to increase its readability, the line plot still suffers from overplotting when visualizing a large number of sensitivity volumes. Therefore, we considered drawing multiple line plots and stacking them in multiple rows. As this limits the amount of available screen space per line plot in the vertical direction, we considered Horizon Graphs as an alternative design to line plots. Horizon Graphs were originally intended to visualize time series~\cite{horizon} and make use of small multiples to show each time series individually.  
For the construction of Horizon Graphs, the data are first shown as an area plot, see Figure~\ref{fig:horizonGraphCreation}a. Then, the area is divided into discrete, equally sized bands which are colored following a continuous color map. We adapt the splitting and coloring scheme by giving the bands a fixed height and, thus, adapting the number of bands to the data instead of fixing the number of bands and adapting their size. For normalized sensitivity values (Sobol indices and $\delta$ sensitivity), we choose a bandwidth of 0.2, which corresponds to a maximum of $5$ bands. For DGSA, we choose a bandwidth of $1$. Note that sensitivity values are always non-negative. Then, we choose a gray color for the first band. This corresponds to non-sensitive values for DGSA and can also approximately be considered non-sensitive for the other sensitivity measures. For the sensitive values, we choose a continuous white-to-red color map such that more reddish colors indicate higher sensitivity values, see Figure~\ref{fig:horizonGraphCreation}b. Next, the bands are collapsed and superimposed to free up vertical space, see Figure~\ref{fig:horizonGraphCreation}c. Horizon Graphs allow for displaying multiple time series without clutter, while allowing for reading off exact values. Due to the coloring, it is also possible to directly spot regions with highest values. However, this visual design needs a lot of vertical space. If the vertical space is limited, e.g., because of the use of multiple coordinated views, the individual plots become rather small, see Figure~\ref{fig:sensitivityVisDesign}b.

To summarize the scalability of our two design choices, we can state that line charts scale well with respect to the required space but suffer from overplotting, while Horizon Graphs produce no overplotting but require much space. We trade off these advantages and drawbacks by combining the two visual encodings in our final design choice:
We render the first $m$ sensitivity volumes using Horizon Graphs and combine the remaining volumes in a line chart visualization, see Figure~\ref{fig:sensitivityVisDesign}c. Thus, we limit the amount of required vertical space and at the same reduce overplotting. The number of Horizon Graphs $m$ can be chosen by the user and interactively changed to adapt the visualization to the data and the available screen space. 
The default order of the sensitivity volumes is the same as above such that the volumes with highest sensitivities are shown in the Horizon Graphs and the presumably less important volumes are summarized in the line plot. Changing the axes order in the PCP is linked with the ordering of the Horizon Graphs and, thus, allows for a user-defined order.
It should be noted that the amount of data can be reduced further by using the same filtering options as presented for the PCP in Section~\ref{sec:pcp}. Moreover, it is possible to interactively brush in the plots and, thus, select single or multiple spatial regions, see the region highlighted with a gray box in Figure~\ref{fig:sensitivityVisDesign}c. The selected voxels are also highlighted in the PCP and their 3D positions are shown in the surface rendering. Even though this spatial sensitivity visualization directly includes spatial information, the linking to the 3D spatial visualization is important to interpret their location and to differentiate features in the data from possible artifacts introduced by the dimensionality reduction.

\section{Parameter dependency visualization}

\begin{figure}
\centering
\includegraphics[width=\linewidth]{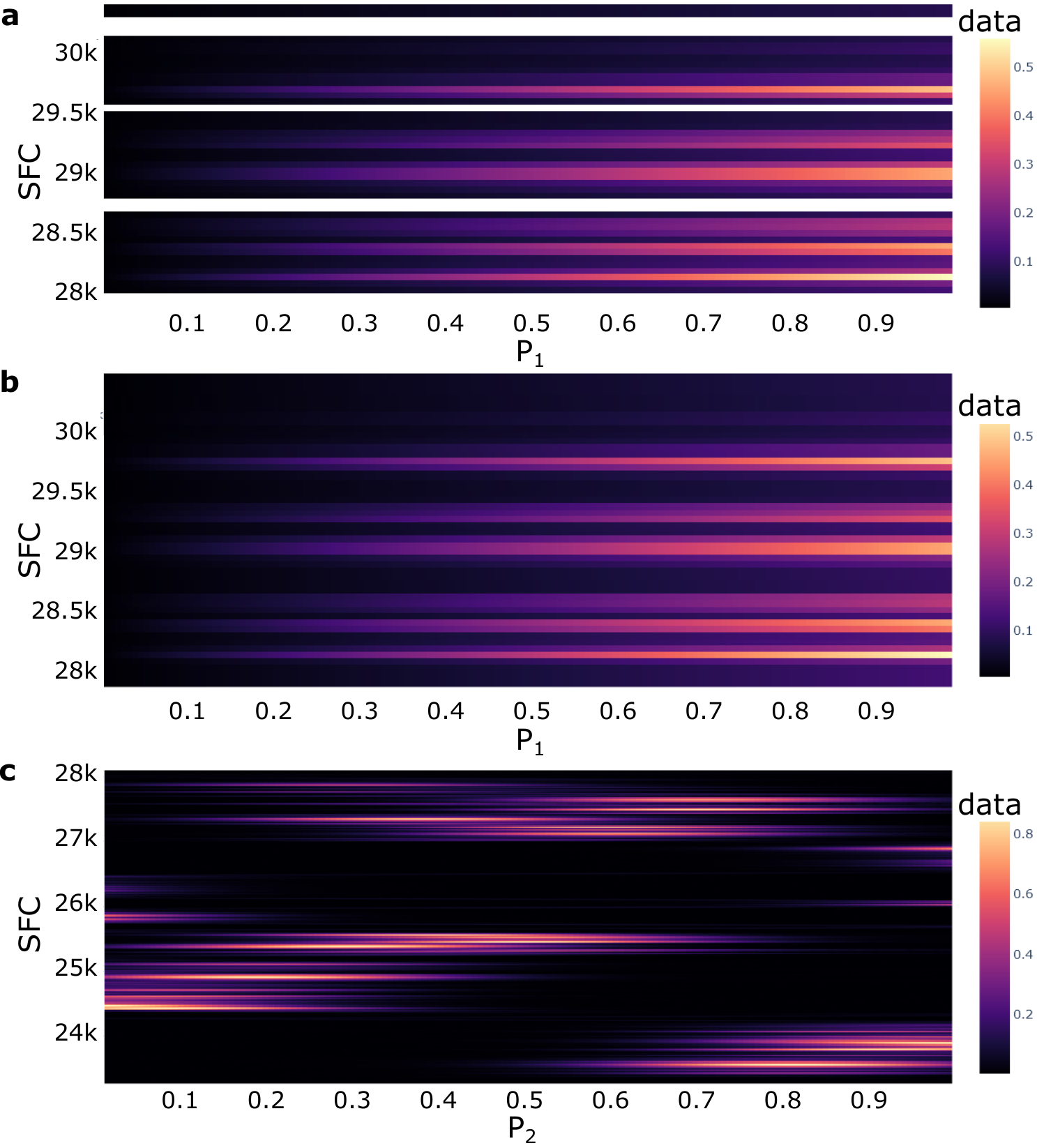}
\caption{Parameter dependency visualization of synthetic dataset. Simulation output is shown as colors in a 2D heatmap spanned by the parameter values (horizontal axis) of parameter $P_1$ (\textbf{a}, \textbf{b}) and parameter $P_2$ (\textbf{c}) and the space-filling curve (vertical axis). \textbf{a} Missing values can be caused by spatial selections and are indicated by gaps, which complicate the interpretation of patterns. \textbf{b} Filling the gaps provides an undisturbed view on occurring patterns like the increase of the values with increasing parameter values. \textbf{c} Showing a different spatial selection over parameter $P_2$ reveals that the spatial regions of high intensity vary with changes in $P_2$.}
\label{fig:parameterDependency}
\end{figure}

After having identified interesting regions that show sensitivities to one or multiple parameters, it is of interest to investigate how the simulation outcome depends on the respective parameter in different spatial regions (T4). We visually encode this information in a 2D heatmap that shows the simulation output over a Cartesian coordinate system, where the horizontal axis represents the parameter values of the selected parameter and the vertical axis represents the spatial dimensions using the SFC as above, see Figure~\ref{fig:parameterDependency}. Even though switching the axis would allow for an easier relation to the sensitivity visualization described above, we decided for this setup, as it displays the change with varying parameter values along the horizontal axis, which has been shown to be more intuitive~\cite{woodin2021conceptual}.
Note that this visualization shows the actual simulation output, while the previously discussed visualizations all relate to sensitivity values. Additionally, the parameter dependency visualization not only aims at investigating changes in the parameter values but also spatial patterns that would not become visible in visualizations that do not preserve the spatial order, such as multiple line charts.
 The simulation output of the respective volume is encoded using matplotlib's perceptually uniform \texttt{magma} color map~\cite{mplMagma}.  Due to the large amounts of data, we use spatial subsampling as above. We aggregate the data by creating a grid and computing the mean of each grid cell. Here, we also aggregate over the other parameters that are not selected to be shown in this visualization. Even though this smooths the data and removes some outliers, overall trends are captured if the grid is chosen fine enough. For the examples shown in this paper, we chose a grid resolution of $150\times 500$. The resolution in the horizontal direction is lower, because there are typically significantly less ensemble members than voxels in the volume and, thus, the sampling in parameter space is less dense.

When aggregating data over the grid cells, some of the grid cells may actually be empty, as there might be gaps without data available, if the parameter space is irregularly sampled or only a subset of the voxels is selected.
Since the user should clearly see if there are data missing, we include the gaps into the visualization, see Figure~\ref{fig:parameterDependency}a. On the other hand, the shown gaps may make interpretations of transitions more difficult.
Thus, the user has the option to switch to a visualization, where the gaps are filled using nearest-neighbor interpolation, see Figure~\ref{fig:parameterDependency}b. Interactively switching between both types of visualization allows the user to obtain an easier understanding of the data while still being aware of missing data and avoiding wrong interpretations.

Our heatmap visualization of parameter dependency allows for observing different characteristics in the data. First, we can spot changes of values in certain spatial regions. For example, in Figure~\ref{fig:parameterDependency}b, we can see an increase with increasing parameter values. While the exact nature of the increase may be easier to interpret in a visualization, where the simulation output is plotted as a graph over the parameter, using the spatial component combined with a color coding of the simulation data allows the user to estimate the relative spatial extent of the feature. Another feature that could not be spotted in visualizations without spatial information is the motion of regions with certain value ranges. An example is shown in Figure~\ref{fig:parameterDependency}c. This observation corresponds to a Gaussian moving across the volume. Here, the feature in the parameter dependency visualization is split into different regions but still clearly visible. The splitting occurred, because the SFC was created by taking sensitivities into account and not the actual data values. The whole region, where the Gaussian passes by during the variation of the respective parameter, is considered sensitive.

\label{sec:parameterDependency}

\section{Results and discussion}
In the following, we validate our approach by applying it to a synthetic dataset, also studying the different algorithmic parameters included in our approach. Afterwards, we present the analysis of real-world ensemble data.

\subsection{Datasets}
\label{sec:datasets}
We create a \textit{synthetic dataset} to verify that our calculation of sensitivity volumes works correctly and our visualizations show the expected features. Therefore, we create an ensemble of $4096$ members with a spatial resolution of $32\times 32\times 32$. The dataset has a 3-dimensional, irregularly sampled parameter space with parameters $P_1$, $P_2$ and $P_3$ that all lie in the range $[0,1]$. Parameter $P_3$ does not influence the result, while the others influence the intensity and position of Gaussian kernels. Each member's scalar field $g(\mathbf{x})$ of the dataset is created using the following function:
\begin{eqnarray*}
g(\mathbf{x}) & = & P_1\cdot f(\mathbf{x};(7,7,7),3) + P_1\cdot P_2\cdot f(\mathbf{x};(10,25,15),3) \\ & & + f(\mathbf{x};(20, 20, 5+P_2\cdot 20),3) + \zeta,
\end{eqnarray*}
where $\zeta$ denotes uniform random noise between $0$ and $0.01$ and $f(\mathbf{x};(x_1, x_2, x_3),\sigma)$ is a 3D Gaussian kernel with a standard deviation of $\sigma$ that is centered at $(x_1, x_2, x_3)$. We refer to the output values as \textit{Output}.

The second dataset is an ensemble of \emph{blood flow simulations} which model the flow through an aneurysm driven by four input parameters. The \emph{viscosity} and the \emph{density} of the blood as well as the maximal flow velocity in the parabolic inlet profile directly influence the flow properties. The fourth parameter is the so-called Smagorinsky constant which is a dimensionless model parameter for the turbulence model. For this paper, we investigate the influence of the input parameters on the flow magnitude. The dataset contains 320 simulation runs with a spatial resolution of $257\times 119\times 128$. However, as the number of samples in each dimension needs to be even for computing the SFC, we resample the dataset to a resolution of $128\times 64\times 64$. We also reduced the resolution to avoid upsampling in the second dimension.

As a third dataset, we investigate \textit{radiofrequency ablation simulations} as a real-world use case~\cite{heimes_studying_2022}. 
The dataset contains temperature fields as the result of radiofrequency ablations in the liver. We use an ensemble of $1,024$ members with a spatial resolution of $92\times92\times92$. The simulation outcome is influenced by the thermal conductivity (\textit{TC}), the blood perfusion rate (\textit{BPR}), the \emph{speed of sound}, the tissue \emph{density}, and the heat capacity (\emph{HC}) of three different tissue types, namely liver (\textit{L}), vessel (\textit{V}) and tumor (\textit{T}). Thus, there are in total $15$ varied parameters. For this dataset, we also include the ensemble's probability of ablation to the PCP, as domain experts rated this information to be highly important for the interpretation of the results.

\subsection{Evaluation of algorithms}
We first evaluate the choice of the sensitivity computation algorithm and then compare the different distance measures for the computation of the SFC.

\subsubsection{Sensitivity Computation}
\label{sec:sensitivityComputationEval}
\begin{figure*}[hbt!]
\centering
\includegraphics[width=\linewidth]{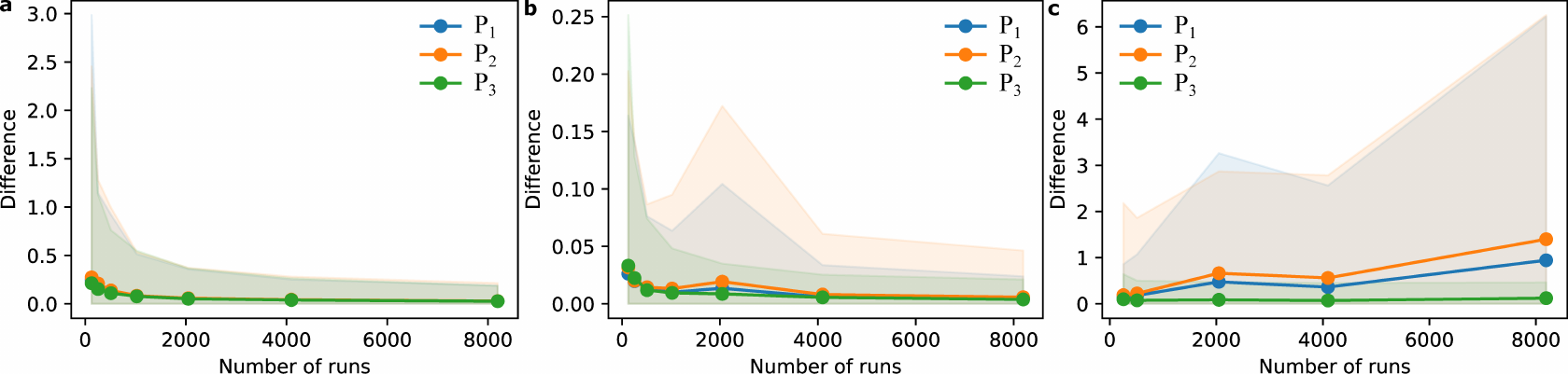}
\caption{
  Comparison of the convergence of Sobol indices (\textbf{a}), $\delta$ sensitivity measure (\textbf{b}), and DGSA (\textbf{c}) with respect to the number of runs. The solid line shows the mean over all voxels of the absolute differences to the previous computation, and the shaded areas show its total range. The synthetic dataset with parameters $P_1$, $P_2$, and $P_3$ is used.
}
\label{fig:convergence}
\end{figure*}

As a first evalation, we compare Sobol indices, $\delta$ sensitivitiy measure, and DGSA. For the comparison, we consider convergence with an increasing number of simulation runs, computations times as well as a visual comparison. For the first two criteria, we use the synthetic dataset and $16$ to $8,192$ parameter-space samples which were created by using Saltelli sampling~\cite{saltelli2008global}. Thus, we obtain a sequence of simulation ensembles with increasing number of parameter-space samples.

\begin{figure}[hbt!]
\centering
\includegraphics[width=\linewidth]{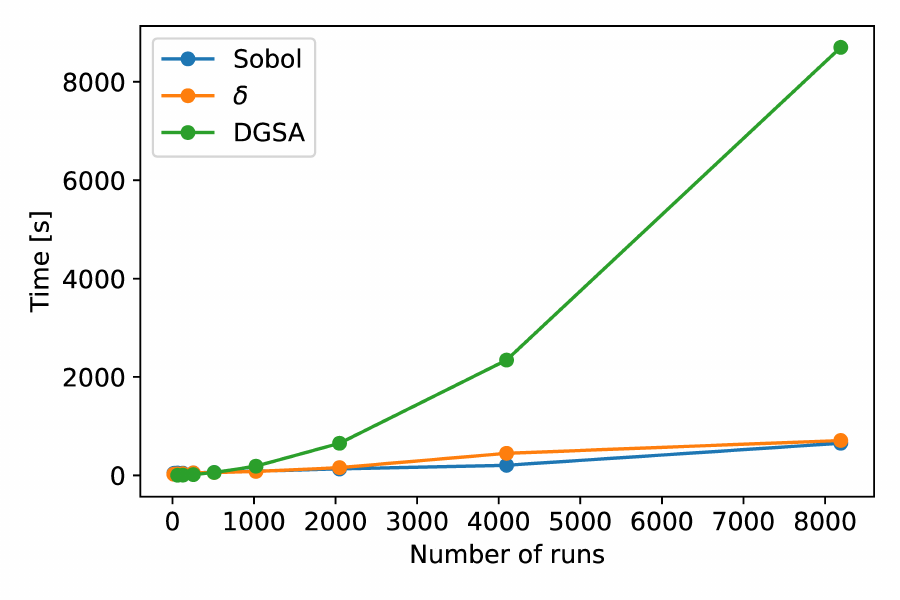}
\caption{The computation times for DGSA increase significantly faster with the number of runs than for the other two sensitivity measures.
}
\label{fig:timings}
\end{figure}

For investigating the convergence, we compute the difference of the sensitivity values for each voxel to the values for the simulation ensemble with less runs, i.e., the preceding ensemble in the generated sequence. These differences are aggregated by computing the mean of the volume. However, we also consider the variation by including the range from the minimal value to the maximal value. The results for the convergence of all three methods are shown in Figure~\ref{fig:convergence}. The result for the Sobol sensitivity indices (see Figure~\ref{fig:convergence}a) reveals a high deviation for less than $1,000$ samples that decreases with an increase of the samples. For the $\delta$ sensitivity measure (see Figure~\ref{fig:convergence}b), the deviation is significantly smaller, also for less runs. However, the decrease is also less clear and for $2,048$ runs, an outlier can be observed. In case of DGSA (see Figure~\ref{fig:convergence}c), no decrease is visible. However, the absolute values cannot be compared to those of the other sensitivity measures. While the others are normalized to the range $[0,1]$, for DGSA a voxel can be considered sensitive to the respective parameter if the sensitivity value exceeds $1$. The timings for computing the sensitivity volumes with the different measures are shown in Figure~\ref{fig:timings}. 
While the computation times of the Sobol indices and the $\delta$ sensitivity measure increase approximately linear with an increase in the number of runs, the computation times required to compute DGSA quickly increase to significantly higher numbers.

\begin{figure*}[hbt!]
\centering
\includegraphics[width=0.85\linewidth]{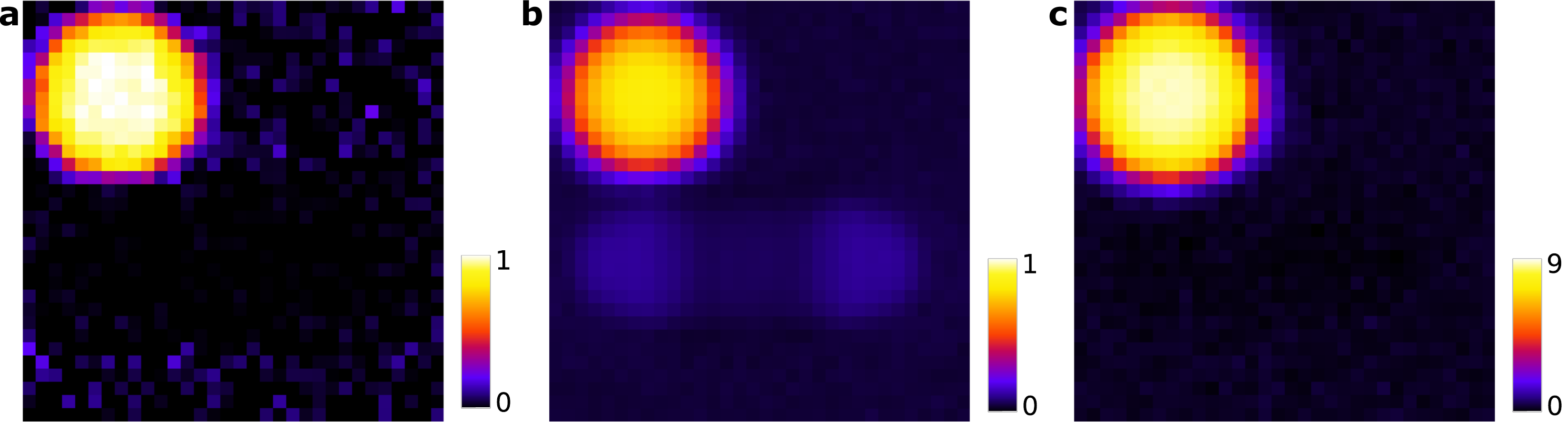}
\caption{
  The sensitivity of the synthetic dataset to parameter $P_1$ computed using Sobol indices (\textbf{a}), $\delta$ sensitivity measure (\textbf{b}), and DGSA (\textbf{c}) is encoded by color in the spatial domain.
}
\label{fig:slicesSynthetic}
\end{figure*}

We also compare the sensitivity measures for synthetic data visually as shown in Figure~\ref{fig:slicesSynthetic}. We see that all sensitivity measures detect the main characteristics. The peak in the upper left is detected by all three sensitivity measures and strongly influenced by parameter $P_1$. The horizontal region in the lower part of the image is correctly detected as not sensitive by Sobol indices and DGSA. However, for the $\delta$ sensitivity measure, we observe slightly increased values on the left and right end of the region. The visualization of the Sobol indices also shows a significant amount of noise, especially in the background. This pattern can be explained by the definition of the Sobol indices, which is based on distributing the variance of the output to the input parameters. As the ensemble members only differ by noise in these regions, this variance cannot be correctly attributed to the input parameters.

\begin{figure}[hbt!]
\centering
\includegraphics[width=\linewidth]{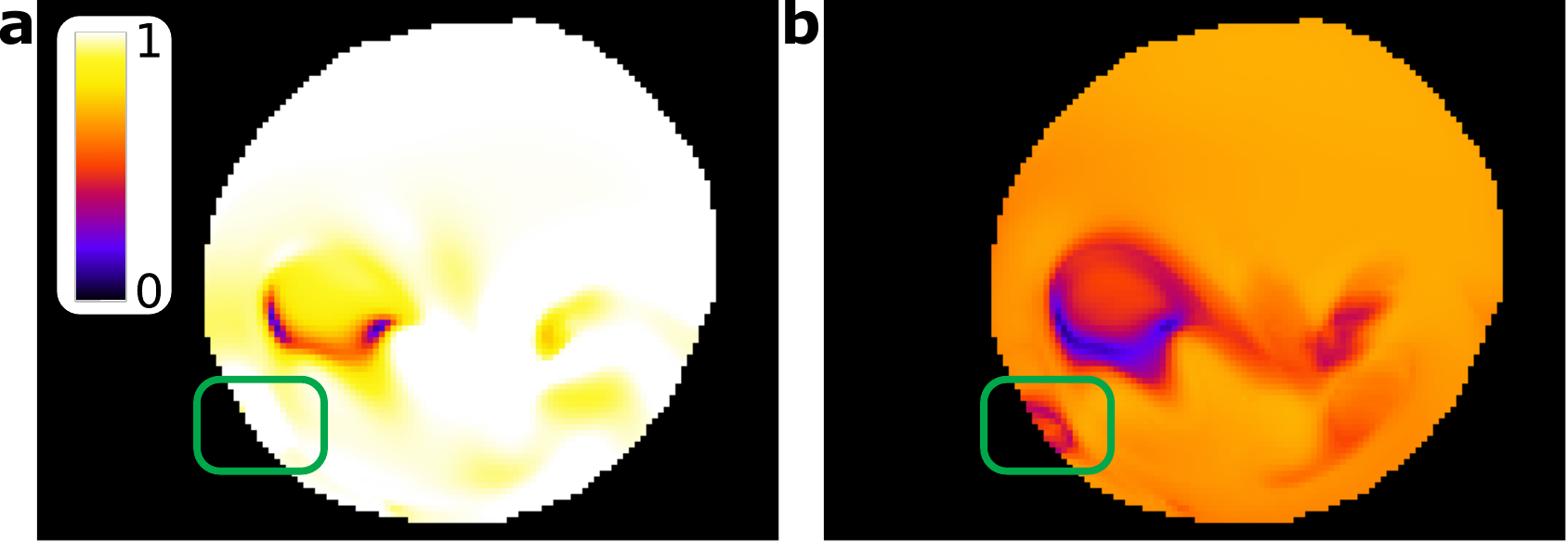}
\caption{
  The sensitivity of the blood flow simulations to the inlet velocity is computed using Sobol indices (\textbf{a}) and $\delta$ sensitivity measure (\textbf{b}). Both images use the same color map. The green frame highlights a region showing a structural difference.
}
\label{fig:slicesAneurysm}
\end{figure}

The sensitivity computations for the blood flow data are shown on a single slice in Figure~\ref{fig:slicesAneurysm}. For the Sobol indices, a significant number of voxels have a sensitivity value outside of the expected range $[0,1]$, which is shown visually as the white areas. This strongly indicates that not enough parameter-space samples were used. We also observe this phenomenon for the ablation dataset (see supplementary material). However, due to the computational costs of creating the simulation results, no more samples are available. This observation agrees with findings in literature that point out the large number of required samples~\cite{saltelli2008global}. Nevertheless, we observe a similar spatial structure for both sensitivity measures. While the shapes of the less sensitive regions vary slightly, the green box indicates a region where a decrease in sensitivity is observed for the $\delta$ sensitivity measure, which is not visible in Sobol indices. One reason for this observation might also be the insufficient numerical accuracy of the Sobol indices, as structural differences in this region are revealed by adapting the color map. However, as these variations lie outside the range of $[0,1]$, they do not allow for meaningful interpretations in Sobol indices.

In summary, we conclude that each of the three sensitivity computation method has advantages and disadvantages. While Sobol indices are commonly used and intuitive to interpret, the number of available samples is often too small in spatial simulations. DGSA produces the smoothest results on the synthetic dataset and can be easily generalized beyond scalar data, but is costly to compute and does not converge if the number of runs is increased. The $\delta$ sensitivity measure produces reasonable results even if some spatial features, such as the horizontal structure in Figure~\ref{fig:slicesSynthetic}b, do not correspond to a sensitive region. However, the sensitivity values are small in this region. Therefore, we use the $\delta$ sensitivity measure for the remainder of the paper.

\subsubsection{Space-filling Curve Computation}
For evaluating the quality of the SFC, we choose the same autocorrelation as defined by Zhou et al.~\cite{zhou2020data}. To investigate the coherency of the data values, we also use the autocorrelations of the data values linearized along the SFC. The autocorrelation is computed for each scalar field individually. The results are averaged to obtain the mean autocorrelation over all fields. For the positional coherency, we use the radial Euclidean distances as proposed by Zhou et al. Thus, we define the function $t(i)=\left\lVert\boldsymbol{p}_i-\boldsymbol{p}_{\mathrm{ref}}\right\rVert$ where $\boldsymbol{p}_i$ is the spatial position of the $i$-th point on the SFC and $\boldsymbol{p}_{\mathrm{ref}}$ denotes the reference point which is chosen as $\boldsymbol{p}_{\mathrm{ref}}=(0,0,0)$.

\begin{figure*}
\centering
\includegraphics[width=\linewidth]{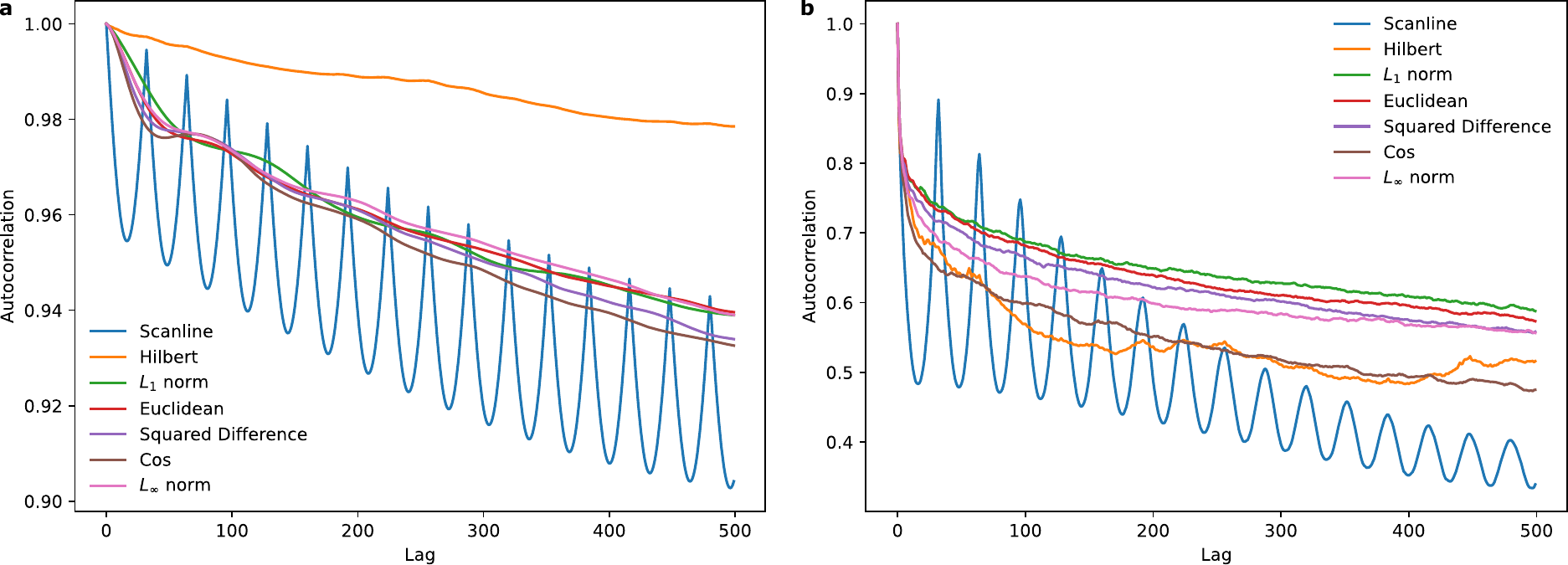}
\caption{Comparison of different space-filling curve algorithms, where the data-driven space-filling curve is tested with different norms for the computation of value coherence. The positional coherency (\textbf{a}) and the value coherency (\textbf{b}) are averaged for the sensitivity indices of the Sobol indices and $\delta$ sensitivity for all three datasets. While the scanline approach performs worst for both measures, the Hilbert curve performs best for the positional coherency, as expected. The data-driven space-filling curve using the $L_1$-norm performs best when considering both measures.}
\label{fig:sfcCorr}
\end{figure*}

For evaluating the five distance measures discussed in Section~\ref{sec:sfc}, we compare the SFCs computed with the different distance measures but also include the scanline algorithm and the Hilbert curve, which both do not consider the underlying data values. To obtain comparable results, all datasets were resampled to a resolution of $32\times 32\times 32$. For each SFC algorithm, we computed the correlations for each dataset and the Sobol as well as $\delta$ sensitivity and aggregated them by computing the average. The results are presented in Figure~\ref{fig:sfcCorr}. As expected, the scanline approach leads to the worst results for both criteria. While the Hilbert curve performs best for the positional coherency (see Figure~\ref{fig:sfcCorr}a), it is outperformed by several of the data-driven approaches for the value coherency (see Figure~\ref{fig:sfcCorr}b). The performance of the data-driven SFCs is comparable for the positional coherency where the $L_\infty$ norm and the $L_1$ norm perform best. However, for the value coherency, the differences among the different distance measures are larger, and the $L_1$ norm performs best, closely followed by the Euclidean distance. As the $L_1$ norm performs well regarding the positional coherency as well as value coherency, we recommend this distance measure and will use it for the remainder of the paper.

\subsection{Use cases}
\label{sec:usecases}
In the following, we validate our approach based on a synthetic dataset and provide two use cases which we discuss with domain experts who were involved in creating the data.

\textbf{Synthetic data.}
We use the synthetic dataset to validate that the desired and expected features in the data are visible in the corresponding visualizations. 

A first overview of the general sensitivity values is provided by the PCP as shown in Figure~\ref{fig:overview}. As the sensitivity values for parameter $P_3$ are very small, it does not influence the result. The other parameters show sensitive regions. This observation agrees with the definition of the dataset. After selecting values with a high sensitivity in parameter $P_1$, the spatial visualization confirms that these values belong to one of the Gaussians. The corresponding parameter dependency visualization is shown in Figure~\ref{fig:parameterDependency}a and b. The linear increase in the simulation output confirms our expectations.

\begin{figure}
    \centering
    \includegraphics[width=\linewidth]{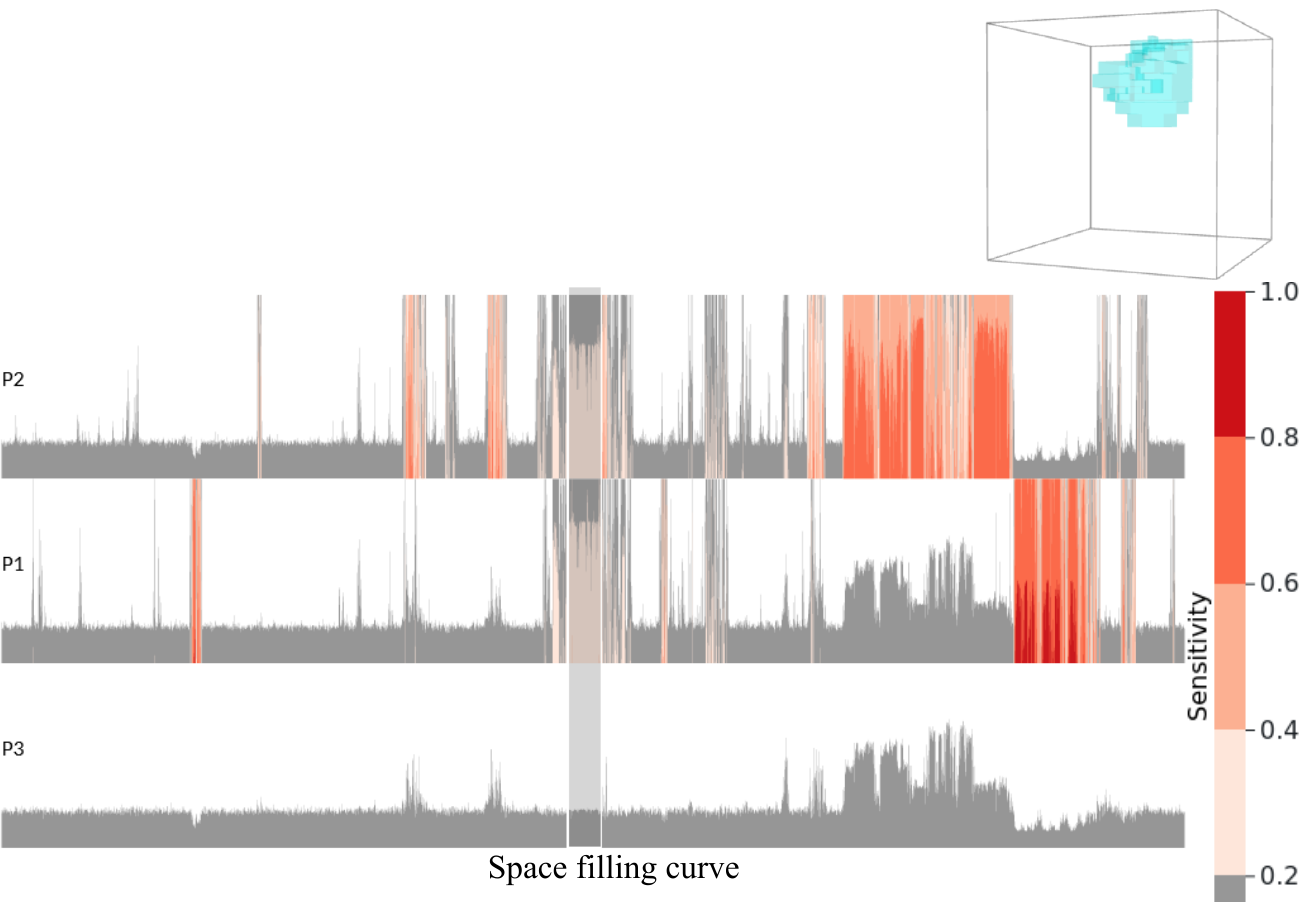}
    \caption{The spatial sensitivity visualization for the synthetic dataset contains regions only sensitive to $P_1$ or $P_2$ as well as regions sensitive to both where the biggest region sensitive to both parameters is selected and shown in the surface visualization.}
    \label{fig:syntheticSpatialSens}
\end{figure}

The spatial variation of the sensitivities can be investigated in the spatial sensitivity visualization as shown in Figure~\ref{fig:syntheticSpatialSens}. While some regions are only sensitive to $P_1$ or $P_2$, some are sensitive to both parameters. Selecting the region sensitive to both parameters shows in the spatial visualization that these values are in the region of the Gaussian whose height scales with the product of $P_1$ and $P_2$. The variation over parameter $P_2$ for the spatial selection that is mainly sensitive to $P_2$ is shown in Figure~\ref{fig:parameterDependency}c. As discussed in Section~\ref{sec:parameterDependency}, it corresponds to a Gaussian moving along the domain. Thus, all findings for the synthetic dataset agree well with its definition.

\begin{figure*}
    \centering
    \includegraphics[width=\linewidth]{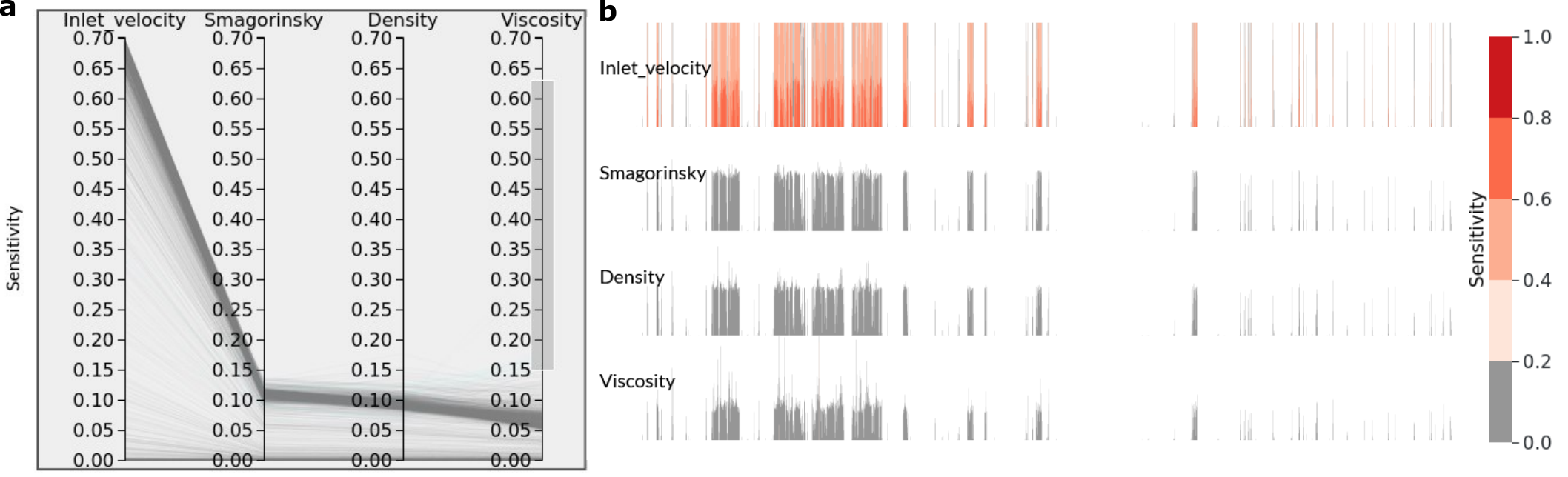}
    \caption{The inlet velocity is the most important parameter for the aneurysm dataset as shown in the PCP (\textbf{a}) and the spatial sensitivity visualization (\textbf{b}). As several voxels of the volume are empty, the spatial sensitivity visualization contains a large number of voxels with a sensitivity of $0$.}
    \label{fig:aneurysmPCPAndSpatial}
\end{figure*}

\begin{figure}
    \centering
    \includegraphics[width=\linewidth]{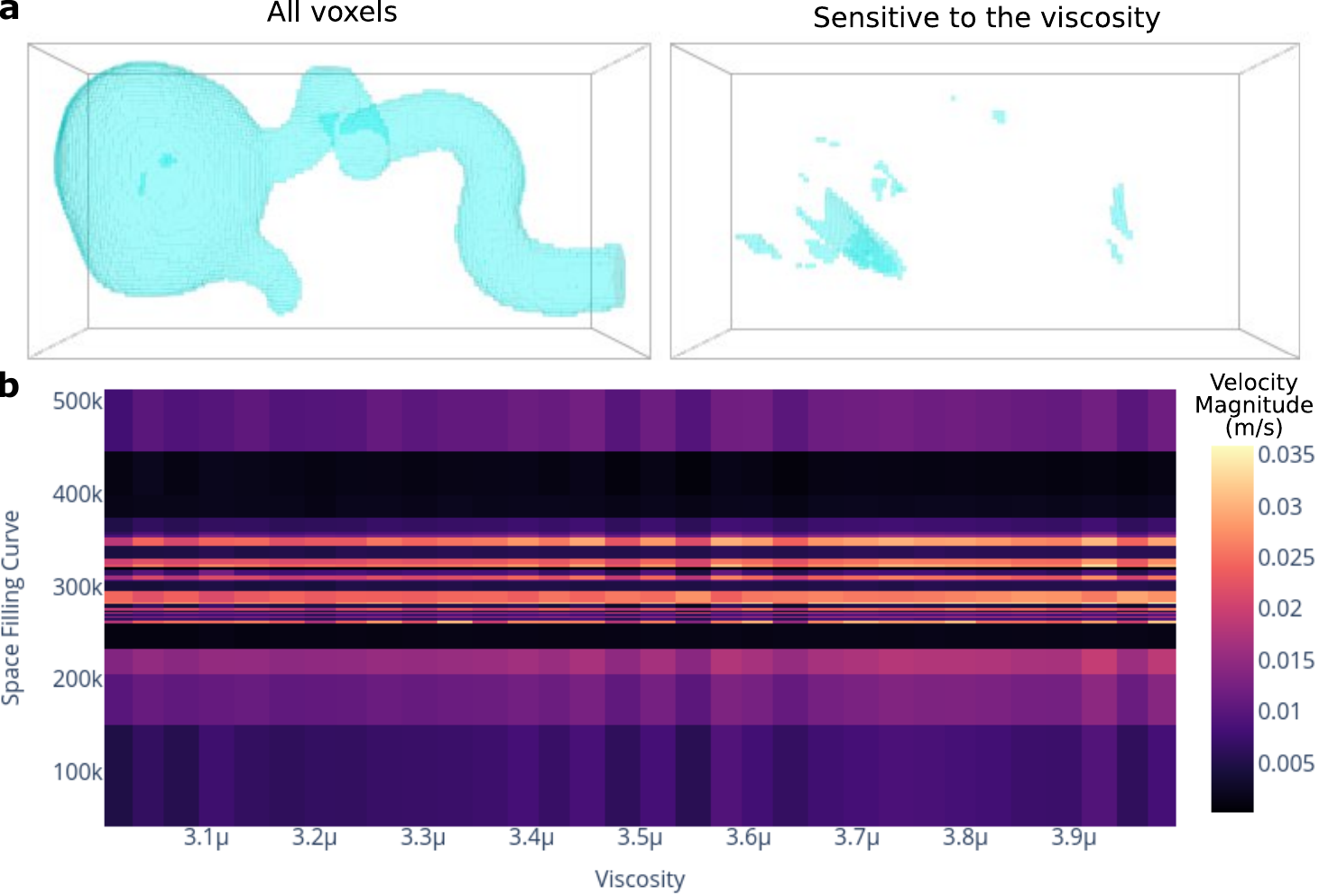}
    \caption{The regions sensitive to the viscosity are scattered across the volume (\textbf{a}). The largest region is located at the inlet of the aneurysm. The parameter dependency visualization (\textbf{b}) shows only a minor increase but the the overall velocity magnitudes in the selected areas are also small.}
    \label{fig:aneurysmViscosity}
\end{figure}

\noindent
\textbf{Blood flow simulations.} The PCP in Figure~\ref{fig:aneurysmPCPAndSpatial}a shows that the inlet velocity is the most influential parameter and the viscosity is the only other parameter to which the output is sensitive in some regions. The influence of the other input parameters can be neglected as the sensitivity values are in the same order of magnitude as those for the irrelevant parameter $P_3$ of the synthetic dataset. The domain expert who created this dataset rated the PCP as especially helpful for obtaining an overview. The spatial sensitivity visualization in Figure~\ref{fig:aneurysmPCPAndSpatial}b shows that the inlet velocity influences almost all voxels belonging to the vessel or the aneurysm (voxels with sensitivity values larger than zero). While the visualization is considered helpful for identifying spatially connected regions of sensitive voxels, the domain expert proposes to remove empty voxels and adapt the data-driven SFC to follow the vessel structure, which would be a solution custom-made for this particular use case. The corresponding values are selected in the PCP to investigate the regions sensitive to density. The PCP also shows a high sensitivity to the inlet velocity for these regions. The 3D rendering shown in Figure~\ref{fig:aneurysmViscosity}a reveals that the corresponding spatial regions are scattered across the domain. The 3D rendering was rated to be essential for spatial context by the domain expert. The largest connected region can be found at the inlet of the aneurysm. This region is related to the inflow region that shows a circulating flow structure and marks the transition between the turbulence in the aneurysm and the laminar flow in the vessel~\cite{bovenkamp4DPhasenkontrastUndMagnetisierungsSattigungstransferMRT2016}.

The parameter dependency visualization as shown in Figure~\ref{fig:aneurysmViscosity}b reveals only a minor increase of the velocity magnitude but the overall magnitude is very small. In the selected regions, the maximum magnitude corresponds to $0.035$~m/s compared to a maximum of $0.1$~m/s over the whole volume which indicates that the slight increase should not be considered significant. These observations agree with the findings of the domain expert, who also sees the possibility of using this visualization for investigating the onset of turbulent flow.

In general, computations of additional simulations for this dataset should focus on investigating the inlet velocity. However, as the viscosity influences the inflow region, which is one of the features of interest in this dataset, it should also be considered. At the same time, the other parameters are less important as their influence is very small. This observation is in agreement with observations on similar simulation setups~\cite{leistikow2020interactive}.

\noindent
\textbf{Radiofrequency ablation data.}
For this dataset with more parameters, it is an important goal to identify the most relevant parameters for the simulation outcome around the tumor region. The tumor should be fully ablated and, at the same time, as little healthy tissue as possible should be ablated. Therefore, a spatial sensitivity analysis is very important because an analysis of the whole dataset might cause misleading results. The different interaction mechanisms used in this analysis and further results on this dataset are shown in the accompanying video. We discussed the analysis with a domain expert who was involved in creating the dataset.

The PCP reveals that for several parameters, no significant sensitivity is detected, which allows for excluding them from the next analysis steps. For example, the speed of sound could be excluded, which was to be expected, as this parameter is not actively used in the simulation model for radiofrequency ablation. The PCP of the most influential parameters is presented in Figure~\ref{fig:teaser}a. The most significant parameter is the liver's blood perfusion rate (BPR), but the BPR of the tumor influences the simulation outcome significantly in several voxels. When tracing the polylines along the axes, one can observe that the voxels sensitive to the tumor's BPR do not agree with those sensitive to the liver's BPR and the tumor's thermal conductivity (TC). It can also be observed that voxels with a high vessel density are also sensitive to the tumor's heat capacity (HC). Thus, the parameters do not only influence the temperature in the corresponding tissue regions but also the surrounding tissue.

The spatial sensitivity visualization shown in Figure~\ref{fig:sensitivityVisDesign}c reveals that the liver's BPC influences most spatial regions, but there are also spatial regions with low sensitivity to this parameter. Instead, these regions' sensitivity to the tumor's TC and HC are increased. Selecting these spatial regions, the 3D visualization shows that the selected voxels are located at the boundary of the liver which is of less interest for the ablation scenario.

\begin{figure}
    \centering
    \includegraphics[width=\linewidth]{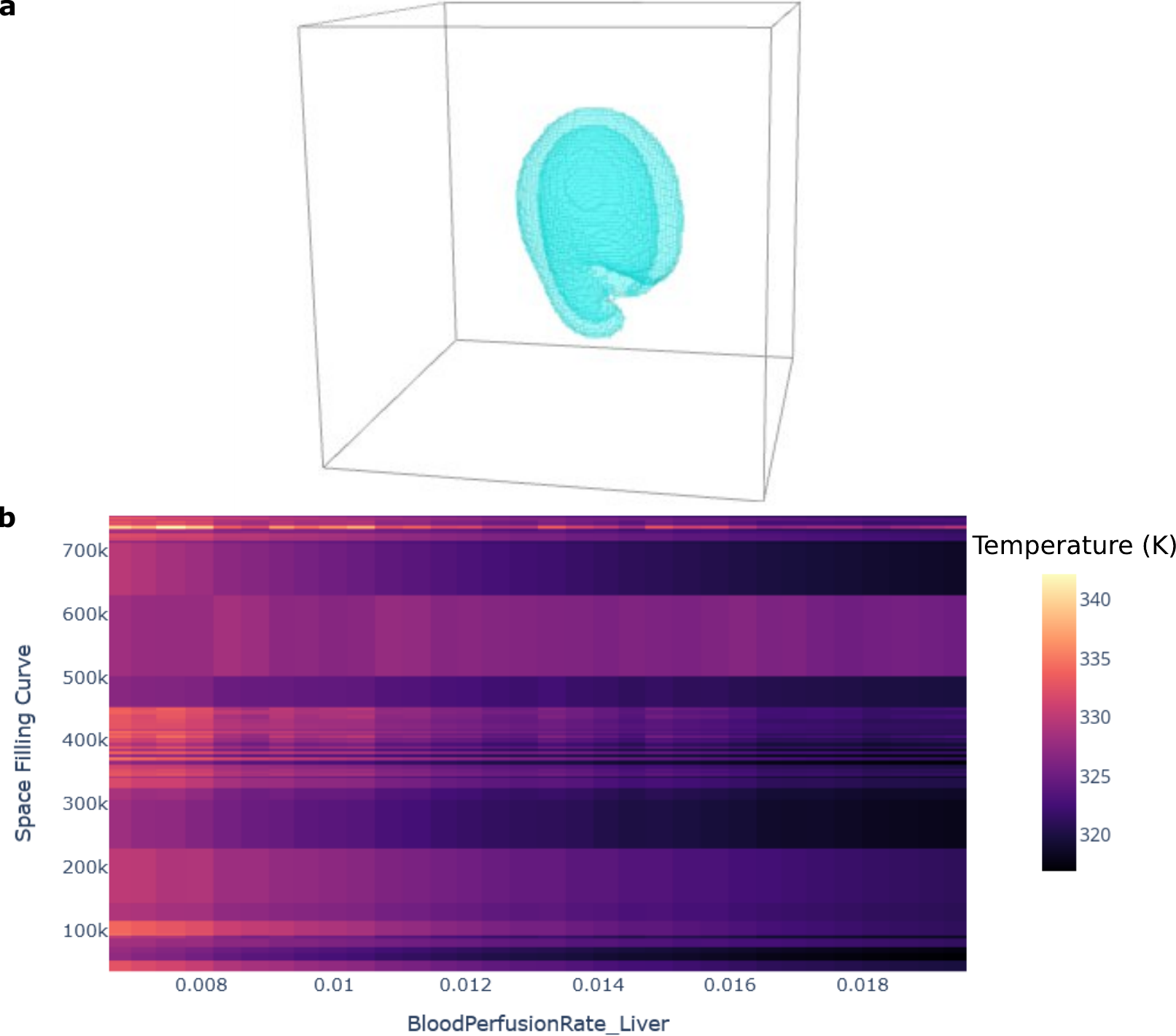}
    \caption{The voxels that are ablated in only some of the ensemble members are located at the boundary of the ablation volume (\textbf{a}). The temperature in this region decreases with an increase of the liver's BPR (\textbf{b}).}
    \label{fig:ablationSelection}
\end{figure}

To limit the analysis to regions of interest, we include the ablation probability into the analysis and the PCP. Based on the ablation threshold of $327.15$~K ($54~^\circ$C), the percentage of ensemble members exceeding this threshold is computed for each voxel. For further analysis, we select an ablation probability $>0$ and $<1$ such that the outcome of the ablation process is uncertain. Figure~\ref{fig:ablationSelection}a shows the selected voxels which are located at the ablation area's boundary. The parameter dependency visualization shown in Figure~\ref{fig:ablationSelection}b shows a decrease in the temperature with increasing BPR of the liver. This observation might be explained by a cooling effect induced by the BPR.  Overall, the findings agree with Heimes et al.~\cite{heimes_studying_2022}. While the domain expert appreciated the possibility of seeing the variation over the parameter, he rated the spatial surface rendering as especially helpful for understanding the important regions. The persistence of ablated regions would be of additional interest. He expressed his interest in investigating more topologically challenging configurations, such as multiple tumors or holes in the ablation volume, which can be caused by the cooling effect of vessels. As our approach is independent of the underlying tissue configuration, it can be applied directly. Additionally, he pointed out that a safety margin around the tumor is relevant for their analysis which can be directly included in the loaded segmentation data. Including explicit comparisons to clinical data and the possibility to investigate interactions between the parameters would be helpful additions to the approach from the domain expert's perspective.

\section{Conclusions and future work}
We propose a methodology for the interactive analysis of spatial sensitivities. We present how spatial sensitivities for volumetric ensembles can be calculated and adapt the calculation of data-driven SFCs to the computation of multi-field sensitivity data. We then propose a visualization approach that supports an interactive analysis by including visualizations of the spatial sensitivity as well as of the simulation's dependence on the different parameters. We discussed our approach with two domain experts who create simulation ensembles. They appreciated our approach and proposed several directions for further applications and future research. 

The scalability with the number of ensemble members is mainly influenced by the computation of the chosen sensitivity measure. However, the computation of the sensitivity values is performed entirely in a preprocessing step, i.e., no further computations are required during the interactive analysis sessions. In contrast to the sensitivity computation, most visualizations are independent of the number of ensemble members. Only the parameter dependency visualization considers the different ensemble members by aggregating over them, which scales well with increasing ensemble sizes. The visual scalability is mainly determined by the number of input parameters and the spatial resolution. The visualizations scale well with the number of parameters. While the sensitivity visualization can show a relatively high number of parameters due to the space-efficient line plot, the PCP limits the scalability. However, the ordering in the PCP allows the user to exclude irrelevant parameters early on in the analysis process. The computation of the sensitivity volumes scales linearly with the number of voxels. As we use a subsampling to support an interactive analysis, the visualizations also scale well with the number of voxels. However, small features might be missed by the sampling for high-resolution datasets.

While we only investigate simulation ensembles that output single scalar fields, the general extension of the approach to time-varying multi-field data, which can be time-varying, multi-field, or both, only requires exchanging the sensitivity computation method. DGSA as proposed by Fenwick et al.~\cite{fenwick2014quantifying} can be easily generalized by exchanging the distance measure.
While the visualizations that purely depend on the sensitivity volumes can be directly applied, future work for finding a parameter dependency visualization for time-varying multi-field  ensembles is necessary.
Further future research directions target a deeper analysis of the interactions between parameters, which was also of interest to one of the domain experts. Besides including interactions between two parameters, higher-order interactions might be interesting in some cases. While most visualizations can be directly applied to higher-order sensitivity indices, a suitable visualization for the qualitative dependencies would need to be developed. We also see great potential to integrate spatial sensitivity analysis with other parameter space analysis approaches. For example, a first step could consist of analyzing the sensitivity and its spatial variations. After identifying interesting features and potentially selecting spatial regions of interest, a partitioning and visualization of the parameter space using hyper-slices~\cite{evers2022multi} could be used to obtain a more comprehensive understanding of the different facets the simulation parameters influence.

\nocite{efron1994introduction}
\nocite{yuan2019research}
\nocite{fisher1958grouping}
\nocite{iwanagaSALibAdvancingAccessibility2022}
\nocite{hermanSALibOpensourcePython2017}
\nocite{saltelliMakingBestUse2002}

\section*{Acknowledgments}
We thank David Sinden for providing domain expert feedback and Sandeep Gyawali, David Sinden, and Tobias Preusser (Fraunhofer MEVIS, Constructor University, Bremen) for providing the radiofrequency ablation dataset. This work was funded by the Deutsche Forschungsgemeinschaft (DFG, German Research Foundation) grants \mbox{260446826},  \mbox{468824876}, and \mbox{431460824 (CRC 1450)}.

\bibliographystyle{IEEEtran}
\bibliography{IEEEabrv,ref.bib}

\newpage

\begin{IEEEbiography}[{\includegraphics[width=1in,height=1.25in,clip,keepaspectratio]{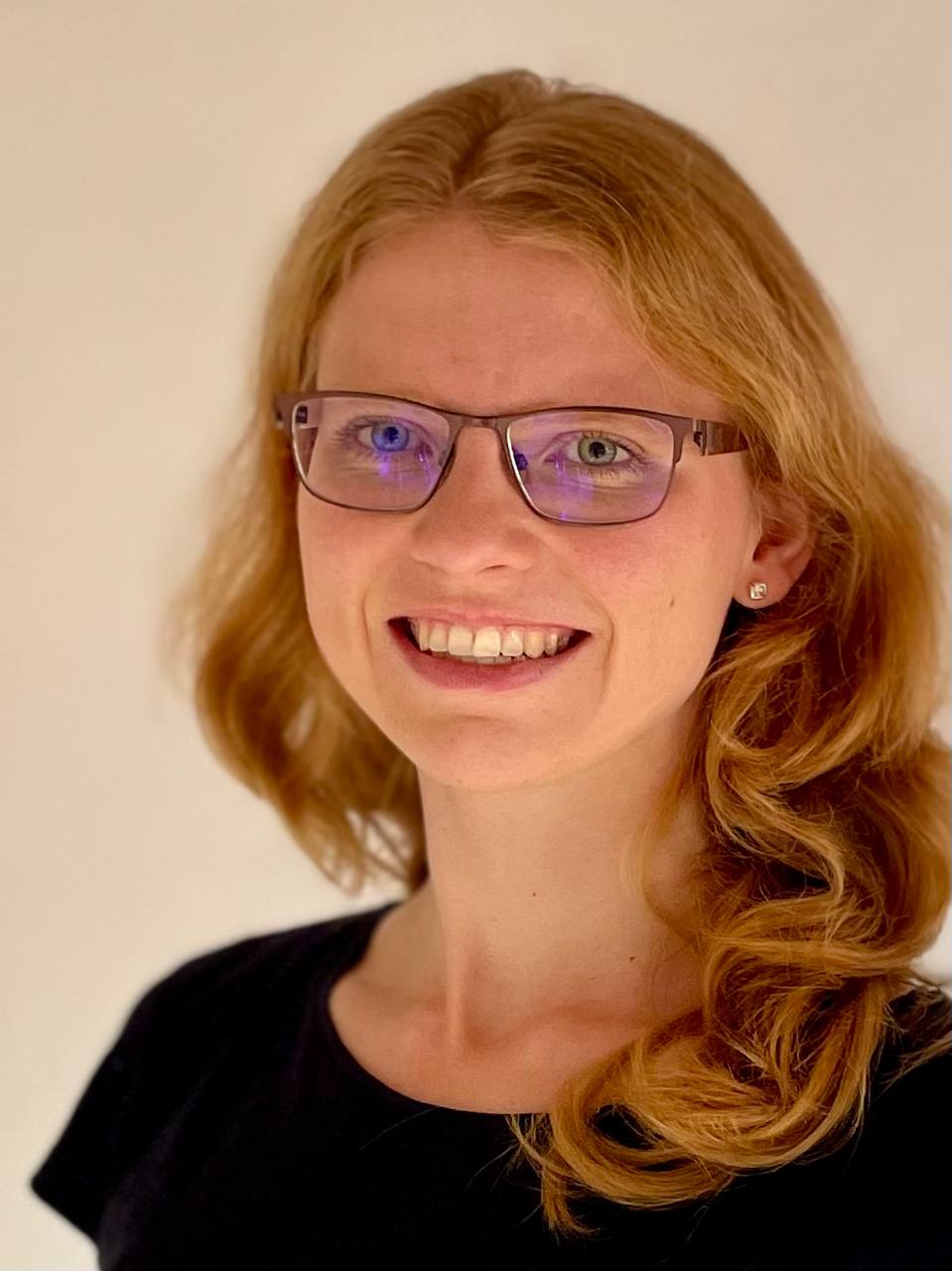}}]{Marina Evers} 
is a postdoctoral researcher at the Visualization Research Center (VISUS) at University of Stuttgart. She received her PhD in Computer Science in 2023 from University of M\"unster, Germany. Her research interests include scientific visualization and visual analytics.
\end{IEEEbiography}

\begin{IEEEbiography}[{\includegraphics[width=1in,height=1.25in,clip,keepaspectratio]{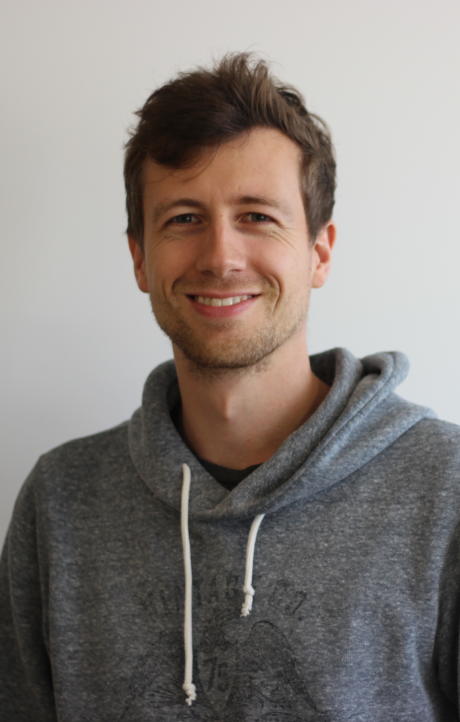}}]{Simon Leistikow} 
is a doctoral researcher in the Visualization and Graphics (VISIX) group at University of M\"unster, Germany, where he received his Master's degree in Computer Science in 2019. His research interests include comparative visual analysis of measured and simulated flow data.
\end{IEEEbiography}

\begin{IEEEbiography}[{\includegraphics[width=1in,height=1.25in,clip,keepaspectratio]{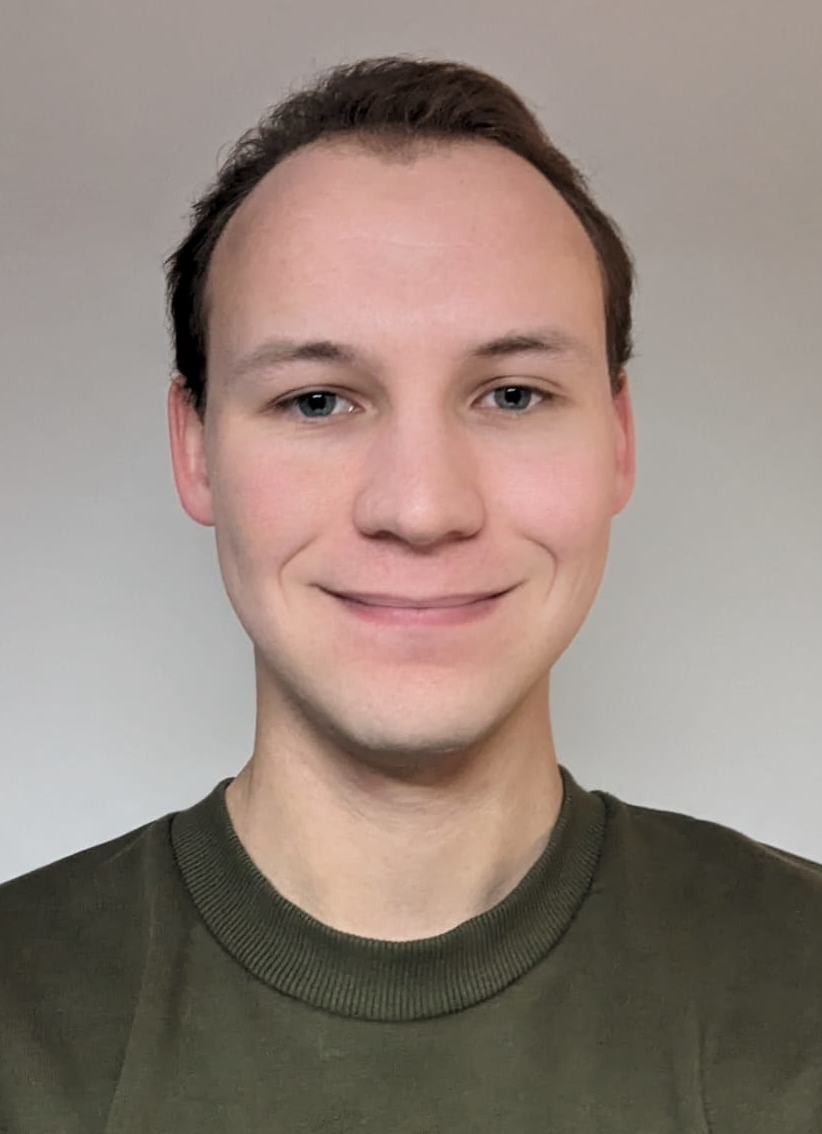}}]{Hennes Rave} 
is a doctoral researcher in the Visualization and Graphics (VISIX) group at University of M\"unster, Germany, where he received his Master's degree in Computer Science in 2021. His research interests include spectral image visualization.
\end{IEEEbiography}

\begin{IEEEbiography}[{\includegraphics[width=1in,height=1.25in,clip,keepaspectratio]{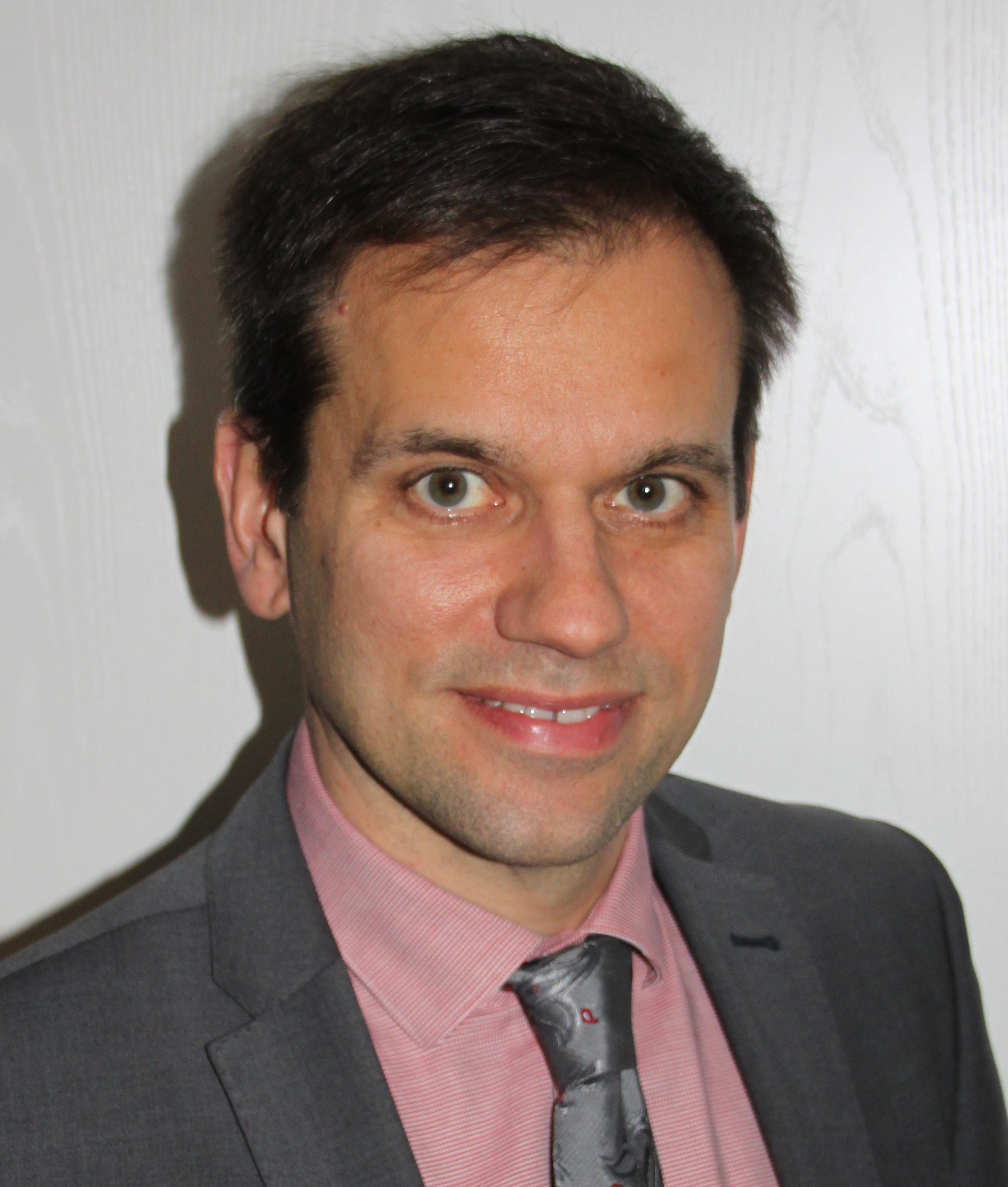}}]{Lars Linsen}
is a Full Professor of Computer Science at University of M\"unster, Germany. He received his academic degrees from the University of Karlsruhe (TH), Germany, including a Ph.D. in Computer Science. Subsequent affiliations were the University of California, Davis, U.S.A., as post-doctoral researcher and lecturer, the University of Greifswald, Germany, as assistant professor, and Jacobs University, Bremen, Germany, as associate and full professor, respectively. His research interests are in interactive visual data analysis.
\end{IEEEbiography}
\vfill

\clearpage

\appendices
\pagestyle{empty}
\section{\label{sec:sensitivity}Sensitivity measures}
In the following, we will briefly explain the computation of the sensitivity indices used in this paper.

\subsection{Sobol Sensitivity Analysis}
The definition of Sobol indices~\cite{sobol2001global, saltelli2008global} is based on the distribution of the total variance $D=\mathrm{Var}[f]$ of function $f$ to the influence of the individual parameters and combinations thereof. Assuming that the parameters are independent and that the function $f$ can be decomposed into subfunctions that only depend on a subset of the parameters, the total variance can be written as the decomposition $D=\sum_{\boldsymbol{\alpha}}V_{\boldsymbol{\alpha}}$ where $V_{\boldsymbol{\alpha}}$ is the variance caused by the parameters $\boldsymbol{\alpha}$. The Sobol indices are then defined as $S_{\boldsymbol{\alpha}}=V_{\boldsymbol{\alpha}}/D$. A common method to compute Sobol indices is based on Monte Carlo methods which use dedicated sampling strategies~\cite{sobol2001global, saltelliMakingBestUse2002} for efficient computation. However, many samples are needed and the number of samples scales poorly with the dimensionality of the input parameter space. For the computations performed in this work, we use the implementation provided by SALib~\cite{hermanSALibOpensourcePython2017, iwanagaSALibAdvancingAccessibility2022} that uses Saltelli sampling.

\subsection{Delta Sensitivity Analysis}
The sensitivity measure $\delta$ was also used by Biswas et al.~\cite{biswas2016visualization} to investigate the sensitivity in weather ensembles. The measure quantifies variations in the simulation output's density function based on variations in the input parameters. The function $g_Y(y)$ denotes the density for the output value $y$ considering the entire output $Y$ while $g_{Y|P_i}(y)$ denotes the density if parameter $P_i$ remains fixed to the value $p_i$. The shift between the two density functions can then be computed as $$s(P_i)=\int |f_{Y}(y)-f_{Y|P_i}(y)|dy\ .$$ The expected value for $s(P_i)$ is calculated by $$E_{P_i}[s(P_i)]=\int f_{P_i}(p_i) s(P_i)dp_i\ ,$$ where $f_{P_i}(p_i)$ denotes the marginal density of $P_i$. Then, the sensitivity measure $\delta_i$ for parameter $P_i$ can be computed as $$\delta_i=\frac{1}{2}E_{P_i}[s(P_i)]\ .$$ As discussed by Borgonovo~\cite{borgonovo2007new}, the value $\delta_i$ lies in the range $[0,1]$, and the sensitivity measure is global, quantitative, and requires no prior assumptions about the model that should be analyzed. We also used the implementation provided by SALib~\cite{hermanSALibOpensourcePython2017, iwanagaSALibAdvancingAccessibility2022} but without the additional computation of Sobol indices.

\subsection{Distance-based generalized sensitivity analysis}
Distance-based generalized sensitivity analysis (DGSA) as originally proposed by Fenwick et al.~\cite{fenwick2014quantifying} uses clustering and cumulative distribution functions. In contrast to other measures such as Sobol indices, DGSA does not require a specific sampling scheme. Additionally, it is only based on distances between the outcome and, thus, can be easily generalized to other data types such as temporal data or non-scalar simulation outcomes.

The first step of DGSA is clustering the ensemble members. As we consider each spatial sample individually, the output in our case consists of scalar values. Therefore, we propose to use Fisher's natural breaks algorithm~\cite{fisher1958grouping} instead of k-medoids as in the original approach. Fisher's natural breaks algorithm is similar to k-means clustering but employs dynamic programming to make use of the 1D structure of the data and can deterministically find a global optimum. Similar to k-means and k-medoids clustering, Fisher's natural breaks algorithm requires the number of clusters $k$ as an input. As this value is strongly data-dependent and might also vary between different spatial locations, we employ an automatic selection of $k$ on each voxel. We propose to use the silhouette coefficient to determine the optimal value of $k$ which is a standard procedure for finding the number of clusters in the $k$-means algorithm~\cite{yuan2019research}. We vary $k$ from $3$ to $10$ clusters, compute the silhouette coefficient for each $k$, and choose the $k$-value with the highest silhouette coefficient. To assure a sufficiently large sample size in each cluster and reduce sensitivity to outliers, we exclude clusters with less than ten ensemble members. Note that the choice of ten members is motivated as a trade-off between choosing a sufficient number of members to compute the \textit{cumulative distribution function} (CDF) and allowing the sensitivity computation also for smaller ensembles.

After clustering the simulation output for each voxel, we continue following the original algorithm. We can calculate a CDF $F(p_i|c_k)$ for each parameter $p_i$ and each cluster $c_k$. 

To quantify the dissimilarity of the distribution functions, we calculate the \textit{distance} $d_{i,k}$ between the distribution function $F(p_i|c_k)$ for a cluster $c_k$ and parameter $p_i$ and the distribution function $F(p_i)$ of the full ensemble for parameter $p_i$. We, therefore, compute
\begin{equation*}
    d_{i,k} = \int_{p_{i, \min}}^{p_{i,\max}} \left| F(p_i|c_k) - F(p_i)\ \mathrm{d}p_i\right|~,
\end{equation*}
where $p_{i, \min}$ and $p_{i, \max}$ are the minimal and maximal parameter value, respectively.

As the interpretation of distance values is difficult, the \textit{statistical significance} of the distance is determined. This is the computationally most expensive step. We use a bootstrapping procedure~\cite{efron1994introduction} as also presented by Fenwick et al.~\cite{fenwick2014quantifying}. To reduce computation costs, we make use of the fact that the threshold distance determined by the bootstrapping algorithm only depends on the number of samples and, thus, can be reused for different voxels in the volumes. We propose to store these data for the whole volume to minimize the number of calculations needed. For the bootstrapping, we use the original algorithm to determine the distance $\hat{d}_{k,i}$ which is the $99\%$ quantile of $B$ randomly selected sets of samples from the original dataset.

When \textit{normalizing} the bootstrapping distance by $d^S_{k,i}=d_{k,i}/\hat{d}_{k,i}$, the simulation outcome can be considered  sensitive to the parameter, if $d^S_{k,i}>1$. As we focus on the overall influence, we use the average normalized bootstrapping distance $s(p_i)=\frac1K \sum_{k=1}^K d^S_{k,i}$ for each parameter $p_i$ and each voxel.

\subsection{Visual Comparison of Sensitivity Computation for Ablation Data}
\begin{figure}[hbt!]
\centering
\includegraphics[width=\linewidth]{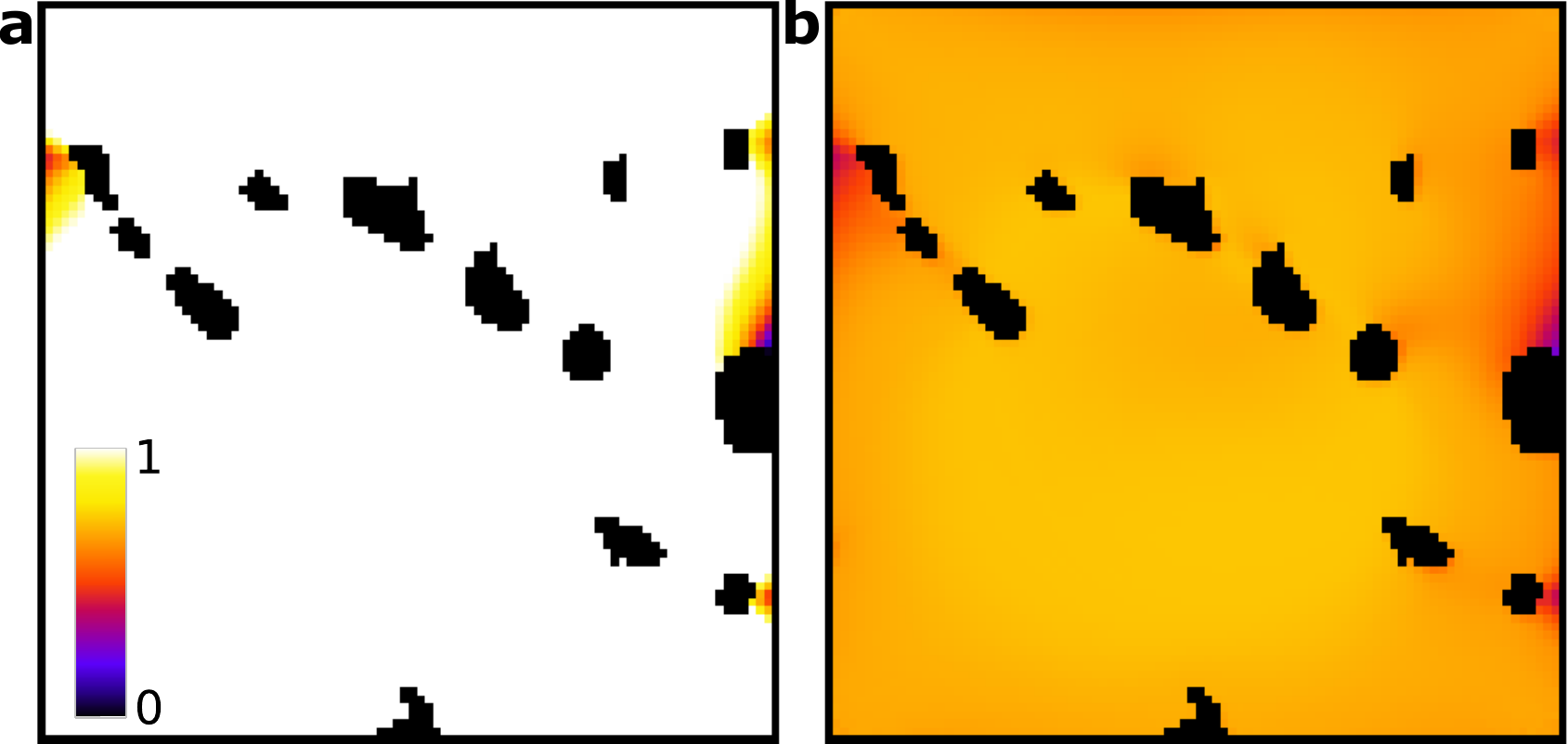}
\caption{
  The sensitivity to the blood perfusion rate of the liver of the ablation dataset is structurally similar for Sobol indices (\textbf{a}) and $\delta$ sensitivity measure (\textbf{b}). Both images use the same color map.
}
\label{fig:slicesAblation}
\end{figure}

Figure~\ref{fig:slicesAblation} provides a comparison of the sensitivities computed by the Sobol indices and the $\delta$ sensitivity measure by color-coding the sensitivity values computed per voxel in a spatial visualization. Similar structures can be observed for both methods.

\vfill

\end{document}